\pdfoutput=1

\documentclass[11pt,twoside,a4paper,cmspaper,final,collab]{cms-tdr}
\def\svnVersion{316224}\def\cmsCernNo{2013-037}\def\cmsCernDate{\today}\def\cmsMessage{Published in Physics Letters B as \href{http://dx.doi.org/10.1016/j.physletb.2015.12.017}{\doi{10.1016/j.physletb.2015.12.017.}}}

\begin{document}\cmsNoteHeader{HIG-14-025}

\hyphenation{had-ron-i-za-tion}
\hyphenation{cal-or-i-me-ter}
\hyphenation{de-vices}
\RCS$Revision: 315496 $
\RCS$HeadURL: svn+ssh://svn.cern.ch/reps/tdr2/papers/HIG-14-025/trunk/HIG-14-025.tex $
\RCS$Id: HIG-14-025.tex 315496 2015-12-17 13:52:36Z klute $

\newlength\cmsFigWidth
\ifthenelse{\boolean{cms@external}}{\setlength\cmsFigWidth{0.98\columnwidth}}{\setlength\cmsFigWidth{0.65\textwidth}}
\ifthenelse{\boolean{cms@external}}{\providecommand{\cmsLeft}{top\xspace}}{\providecommand{\cmsLeft}{left\xspace}}
\ifthenelse{\boolean{cms@external}}{\providecommand{\cmsRight}{bottom\xspace}}{\providecommand{\cmsRight}{right\xspace}}
\newcommand{\vmet}{\ensuremath{\vec{E}_\mathrm{T}}^{\text{miss}}\xspace}
\newcommand{\vg}{\ensuremath{\vec{E}^{\gamma}_\mathrm{T}}\xspace}
\newcommand{\mH}{\ensuremath{m_{\PH}}\xspace}
\newcommand{\V}{\ensuremath{\mathrm{V}}\xspace}
\newcommand{\delphill}{\ensuremath{\Delta\phi_{\ell\ell}}\xspace}
\newcommand{\etg}{\ensuremath{\et^\gamma}\xspace}
\newcommand{\dyll}{\ensuremath{\Z/\gamma^*\to\ell^+\ell^-}\xspace}
\newcommand{\mT}{\ensuremath{m_{\mathrm{T}}^{\ell\ell\MET}}}
\newcommand{\MT}{\ensuremath{m_{\mathrm{T}}^{\gamma\MET}}}
\newcommand{\Zjets}{\ensuremath{\Z+\text{jets}}\xspace}
\newcommand{\mll}{\ensuremath{m_{\ell\ell}}\xspace}

\providecommand{\NA}{---}

\cmsNoteHeader{HIG-14-025}
\title{Search for exotic decays of a Higgs boson into undetectable particles and one or more photons}

\date{\today}

\abstract{
A search is presented for exotic decays of a Higgs boson into undetectable particles
and one or two isolated photons  in pp collisions at a center-of-mass
energy of 8\TeV. The data correspond to an
integrated luminosity of up to 19.4\fbinv collected with
the CMS detector at the LHC. Higgs bosons produced in gluon-gluon fusion and in association
with a Z boson are investigated, using models in which the Higgs boson decays into a
gravitino and a neutralino or a pair of neutralinos, followed by the decay of the neutralino to a
gravitino and a photon. The selected events are consistent with the background-only hypothesis,
and limits are placed on the product of cross sections and
branching fractions.
Assuming a standard model Higgs boson production cross-section, a 95\% confidence level
upper limit is set on the branching fraction of a 125\GeV Higgs boson
decaying into undetectable particles and one or two isolated photons as a
function of the neutralino mass. For this class of models and neutralino masses from 1 to 120\GeV an upper limit in the range of 7 to 13\% is obtained. Further results are given as a function of the neutralino lifetime, and also for a range of Higgs boson masses.
}

\hypersetup{%
pdfauthor={CMS Collaboration},%
pdftitle={Search for exotic decays of a Higgs boson into undetectable particles and one or more photons},%
pdfsubject={CMS},%
pdfkeywords={CMS, physics, Higgs}}

\maketitle
\section{Introduction}
The detailed studies of the properties of the observed Higgs
boson~\cite{AtlasPaperCombination, CMSPaperCombination,
  CMSPaperCombination2} are key components of the LHC physics
program. In the standard model (SM) and for a given mass of
the Higgs boson, all properties of the Higgs boson are
predicted. Physics beyond the SM (BSM) might lead to deviations from
these predictions. Thus far, measurements of the Higgs
bosons couplings to fermions and bosons and of the tensor structure of the
Higgs boson interaction with electroweak gauge bosons show no
significant deviations~\cite{Khachatryan:2014jba,Khachatryan:2014kca} with respect to SM expectations.

Measurements of Higgs boson couplings performed for visible decay
modes provide constraints on partial decay widths of the Higgs
boson to BSM particles. Assuming that the couplings of the Higgs boson
to $\PW$ and $\cPZ$ bosons are smaller than
the SM values, this indirect method provides an upper limit on the
branching fraction of the 125\GeV Higgs boson to BSM particles of 57\% at a 95\% confidence level (CL)~\cite{Khachatryan:2014jba, LHCHXSWG1}.
An explicit search for BSM Higgs boson decays presents an alternative opportunity for
the discovery of BSM physics.
The observation of a sizable decay branching fraction of the Higgs boson to undetected (\eg invisible or largely invisible) final states would be a clear sign of BSM physics and could provide a window on dark matter~\cite{ZHTheory, Martin:1999qf, Bai:2011wz, Gori}.

\begin{figure*}[htb]
\centering
\includegraphics[width=0.45\textwidth]{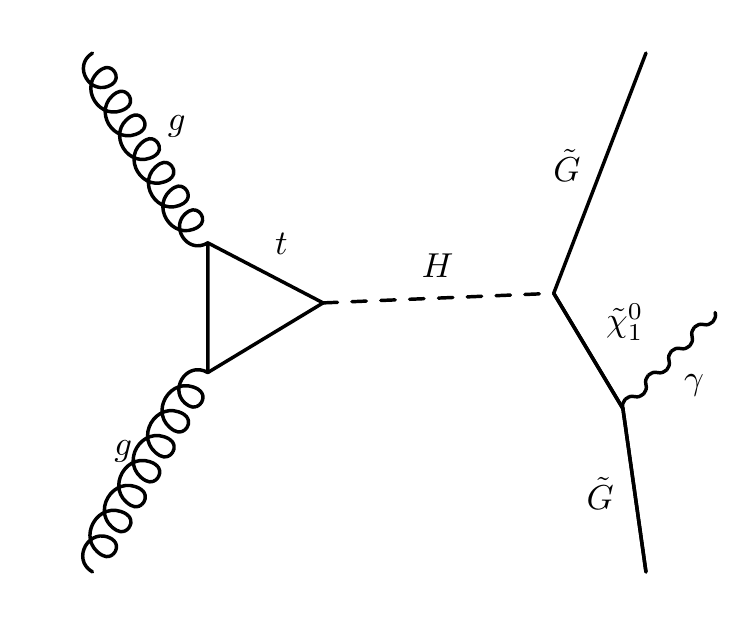}
\includegraphics[width=0.45\textwidth]{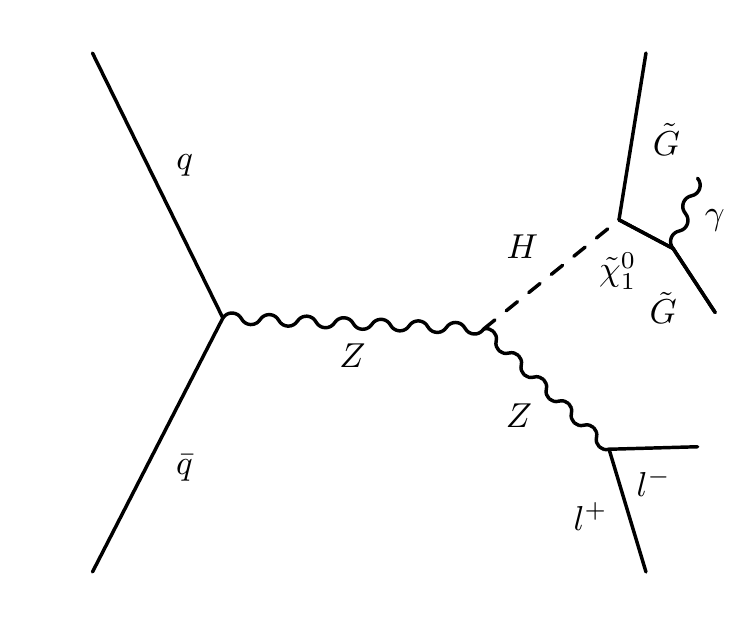}
\caption{Feynman diagrams for the $\PH\to \text{undetectable}+\gamma$
  final state produced via $\Pg\Pg\PH$ (left) and $\cPZ\PH$ (right).}
\label{fig:exo-higgs-decay}
\end{figure*}

Several BSM models predict Higgs boson decays to undetectable particles and photons. In certain
low-scale supersymmetry (SUSY) models, the Higgs bosons are allowed to decay into a gravitino (\PXXSG) and a neutralino
($\PSGczDo$) or a pair of neutralinos~\cite{Djouadi1997243,Petersson:2012dp}.
The neutralino then decays into a photon and a gravitino, the lightest
supersymmetric particle and dark matter candidate. Figure~\ref{fig:exo-higgs-decay} shows
Feynman diagrams for such decay chains of the Higgs boson ($\PH$) produced by gluon-gluon fusion
($\Pg\Pg\PH$) or in association with a $\cPZ$ boson decaying to charged leptons ($\cPZ\PH$).

As the gravitino in these models has a negligible
mass~\cite{Djouadi1997243,  Petersson:2012dp}, the remaining parameter
is the neutralino mass. If its mass is in the range $m_\PH/2 <
m_{\PSGczDo} < m_\PH$, with $m_\PH = 125\GeV$ the mass of the
observed Higgs boson, the branching fraction $\mathcal{B} (\PH
\to \PSGczDo \PXXSG\to \gamma \PXXSG\PXXSG)$ can be
large. For $m_{\PSGczDo} < m_\PH/2$, the decay $\PH \to
\PSGczDo\PSGczDo \to \gamma \gamma \PXXSG\PXXSG$ is expected to dominate.
The same discussion can be applied to heavy neutral Higgs bosons with
masses larger than $125\GeV$.
The lifetime of the neutralino can be finite in some classes of BSM
scenarios, leading to the production of one or more photons displaced
from the primary interaction.

In the SM, the signature equivalent to the signal arises when the Higgs boson decays as
$\PH\to \cPZ\gamma \to \nu\bar{\nu}\gamma$ with a
branching fraction of $3 \times 10^{-4}$. The decay $\PH \to
\cPZ\gamma$ has been studied in $\cPZ \to \Pep\Pem$ and $\cPZ \to \PGmp\PGmm$ final states. Upper
limits on the product of the cross section and branching fraction of
about a factor of ten larger than the SM expectation have been set at the
95\%~CL~\cite{cms-hig-13-006,atlas-hig-14-zgamma}. With the available dataset
the search presented is not sensitive to this decay, but it is sensitive to enhancements
in the Higgs boson decay rates to undetectable particles and photons arising from BSM physics.

Various background processes lead to the signal signatures and are
estimated from simulation or from control samples in data. The dominant
background processes are from $\gamma+$jets events and diboson events in
the $\Pg\Pg\PH$ and $\cPZ\PH$ search, respectively. Details of the background
estimation techniques are discussed in Section~\ref{sec:backgrounds}.
The strength of the $\cPZ\PH$ channel analysis is an almost background-free selection leading to a larger sensitivity in the model-dependent interpretation.
While both the $\Pg\Pg\PH$ and the $\cPZ\PH$ channels provide sensitivity to BSM Higgs boson signatures, the $\Pg\Pg\PH$ channel allows a model-independent interpretation of the results.

This analysis presents a first search for decays of a scalar boson to
undetectable particles and one or two isolated photons. The scalar boson
is produced in $\Pg\Pg\PH$ or in $\cPZ\PH$.
The data used correspond to an integrated luminosity of up to $19.4\pm0.5$\fbinv at a
center-of-mass energy of $\sqrt{s} = 8\TeV$ in 2012 collected with the CMS
detector at the CERN LHC.

The results of the search are presented in terms of the low-scale SUSY
breaking model for $m_\PH = 125\GeV$ and $m_{\PSGczDo}$ between
$1\GeV$ and $120\GeV$, and for $m_\PH$ between $125\GeV$ and
$400\GeV$ for the example case where $m_{\PSGczDo} = m_\PH - 30\GeV$. The effect of a
finite $\PSGczDo$ lifetime ($\tau_{\PSGczDo}$) is studied for the example case where
$m_\PH = 125\GeV$ and $m_{\PSGczDo} = 95\GeV$.

\section{The CMS experiment}
The CMS detector, definitions of angular and spatial coordinates,
and its performance can be found in Ref.~\cite{CMSdetector}.
The central feature of the CMS apparatus is a superconducting
solenoid, of 6\unit{m} internal diameter, providing a magnetic field
of 3.8\unit{T}. The field volume contains a silicon pixel and strip
tracker, a crystal electromagnetic calorimeter (ECAL), and a brass and scintillator hadron calorimeter. Muons are measured in
gas-ionization detectors embedded in the steel flux-return yoke of the
magnet. The first level of the CMS trigger system, composed of specialized
hardware processors, is designed to select the most interesting events
within 3$\mus$, using information from the calorimeters and muon
detectors. A high-level trigger processor farm is used to reduce the
rate to a few hundred events per second before data storage.

A particle-flow
algorithm~\cite{CMS-PAS-PFT-09-001,CMS-PAS-PFT-10-001} is used to reconstruct
all observable particles in the event. The algorithm combines all subdetector information
to reconstruct individual particles and identify them as charged hadrons,
neutral hadrons, photons, and leptons.
The missing transverse energy vector $\vmet$ is defined as
the negative vector sum of the transverse momenta of all reconstructed
particles (charged or neutral) in the event, with $\MET = \abs{\vmet}$.
Jets are reconstructed using the anti-\kt
clustering algorithm~\cite{antikt}  with a distance parameter of $R=0.5$,
as implemented in the \FASTJET package~\cite{Cacciari:fastjet1,Cacciari:fastjet2}.
A multivariate selection is applied to separate jets from the primary interaction
and those reconstructed due to energy deposits associated with pileup
interactions~\cite{jetIdPAS}. The discrimination is based on the differences in
the jet shapes, on the relative multiplicity of charged and neutral components,
and on the different fraction of transverse momentum which is carried by the
hardest components. Photon identification requirements and other
procedures used in selecting events can be found in Section~\ref{sec:selection}.

\section{Data and simulation events}~\label{sec:data_sim}
In the search for Higgs bosons produced in $\Pg\Pg\PH$, the trigger
system requires the presence of one high transverse energy ($\etg$) photon candidate and
significant $\MET$. The
presence of a photon candidate with $\etg > 30\GeV$ is required within the
ECAL barrel region ($\abs{\eta^{\gamma}}< 1.44$).
At the trigger level $\MET$ is calculated
from calorimeter information, and is not corrected for muons. A
selection requirement of $\MET  >$ 25\GeV is applied.
The efficiency of the trigger is monitored and measured with two control triggers
for the photon and the $\MET$ trigger requirement.
The data recorded with this trigger correspond to an integrated luminosity of 7.4\fbinv and were part
of the CMS "data parking" program implemented for the last part of the
data taking at $\sqrt{s} = 8$\TeV in 2012.
In that program, CMS recorded additional data with relaxed trigger requirements,
planning for a delayed offline reconstruction in 2013 after the completion of the LHC Run 1.

For the search for Higgs bosons produced in $\cPZ\PH$,
collision events were collected using single-electron and single-muon
triggers which require the presence of an isolated lepton with $\pt$ in
excess of $27\GeV$ and $24\GeV$, respectively. Also a dilepton
trigger was used, requiring two leptons with $\pt$ thresholds of
$17\GeV$ and $8\GeV$. The luminosity integrated with these triggers at $\sqrt{s} = 8$\TeV is 19.4\fbinv.

Several Monte Carlo (MC) event generators are used to simulate signal and background processes. The
simulated samples are used to optimize the event selection, evaluate selection efficiencies and
systematic uncertainties, and compute expected event yields. In all cases the MC samples
are reweighted to match the trigger efficiency measured in data.

The $\V\gamma$, $\PW\cPZ$, $\cPZ\cPZ$, $\V\V\V$ (where $\V$ represents $\PW$ or $\cPZ$ bosons), Drell--Yan (DY) production of
$\qqbar \to \cPZ/\gamma^*$, and $\qqbar \to\PWp\PWm$ processes are generated with the \MADGRAPH 5.1 event
generator~\cite{madgraph} at leading-order (LO), the $\Pg\Pg \to \PWp\PWm$
process is generated with the LO event generator \textsc{gg2ww} 3.1~\cite{ggww}, and the $\ttbar$ and $\PQt\PW$ processes are
generated with $\POWHEG$ 1.0 at next-to-leading-order (NLO). The
signal samples are also produced with \MADGRAPH. The cross sections at
NLO or higher orders if available are used for a given process to
renormalise the MC event generators. All processes are interfaced to
the \PYTHIA 6.4 generator~\cite{pythia} for parton shower and hadronization.

The CTEQ6L set of parton distribution functions (PDF)~\cite{cteq66} is used for LO
generators, while the CT10~\cite{Lai:2010vv} PDF set is used for NLO generators.  For
all processes, the detector response is simulated with a detailed description of the CMS detector,
based on the \GEANTfour package~\cite{Agostinelli:2002hh}.  Additional $\Pp\Pp$ interactions
overlapping the event of interest in data, denoted as pileup events,
are accounted for by simulating $\Pp\Pp$ interactions with the \PYTHIA generator and
adding them to each MC sample. The MC samples are tuned to reproduce the distribution in
the number of pileup events in data. The average number of pileup events is about
26 for the collected data used in the $\Pg\Pg\PH$ channel, and is about
21 for the collected data used in the $\cPZ\PH$ channel.

\section{Event selection} \label{sec:selection}

Two strategies are followed to isolate the Higgs boson events produced by
$\Pg\Pg\PH$ and by $\cPZ\PH$ from the background
processes. The signal cross sections are several orders of magnitude smaller than
the major reducible background processes, whose contributions are greatly reduced using
the event selections described in the following sub-sections.

\subsection{Event selection in the \texorpdfstring{$\Pg\Pg\PH$}{ggH} channel}

In the $\Pg\Pg\PH$ channel, each selected event is required to have at least one photon candidate
with $\etg>45\GeV$ and $\abs{\eta^\gamma}<1.44$ using a cut-based selection~\cite{Khachatryan:2015iwa,Khachatryan:2014rwa}.
To reduce the SM backgrounds arising from the leptonic decays of $\PW$ and
$\cPZ$ bosons, a lepton veto is applied. Events are rejected if they have one
or more electrons fulfilling a loose identification
requirement~\cite{Khachatryan:2015hwa} and $\pt^{\Pe} > 10\GeV$ and
$\abs{\eta^{\Pe}} < 2.5$, excluding the transition region of
$1.44 < \abs{\eta^{\Pe}} \leq 1.57$ since the reconstruction of an electron object
in this region is not optimal. Similarly, events containing muon candidates with
$\pt^{\mu} > 10\GeV$, $\abs{\eta^{\mu}} < 2.1$, and well separated from the photon candidate requiring
$\Delta R(\gamma,\mu) = \sqrt{\smash[b]{(\Delta\eta)^2 + (\Delta\phi)^2}} > 0.3$
(where the $\phi$ is azimuthal angle in radians), are rejected.
In addition to the selection requirements described above, the $\MET$ is required to be greater than
$40\GeV$. This level of selection is referred to as the
preselection. Additional selection criteria are applied to search for new physics
in either a quasi model-independent way or optimized for a SUSY benchmark model.
In this channel jets can arise from initial-state radiation.
For both search strategies jets are required to have $\pt^{\ensuremath{j}} > 30\GeV$
and $\abs{\eta^j} < 2.4$. These jets must not overlap with the
photon candidate below $\Delta R(\gamma,\text{jet}) < 0.5$.

In the model-independent analysis, events with two or more jets are
rejected. For events with one jet the azimuthal angle between the photon and
the jet ($\Delta \phi(\gamma,\text{jet}))$ is required to be smaller than
2.5. This selection requirement rejects the dominant $\Pgg$+jet
background, where the photon and the jet tend to be back to back in the
transverse plane.

In the model-dependent analysis developed for SUSY scenarios, no requirement is applied on jet multiplicity.
In order to minimize the contribution from processes such as
$\gamma+$jets and multijet events, two methods are used
for identifying events with mismeasured $\MET$. The $\MET$ significance
method~\cite{Chatrchyan:2011tn} takes account of reconstructed objects for each event and their
known resolutions to compute an event-by-event estimate of the likelihood that the observed $\MET$
is consistent with zero. In addition, a minimization
method~\cite{Khachatryan:2014rwa} constructs a $\chi^2$
function of the form

\begin{equation}\label{eq:1ab}
    \chi^2 = \sum_{i=\text{objects}} \left(
             \frac{(\PT^{\text{reco}})_i-(\widetilde{p}_{\mathrm{T}})_i}{(\sigma_{\PT})_i} \right)^2 +
             \left( \frac{\widetilde{E}_x^{\text{miss}}}{\sigma_{{E}_x^{\text{miss}}}}\right)^2 + \left(
             \frac{\widetilde{E}_y^{\text{miss}}}{\sigma_{{\mathrm{E}}_y^{\text{miss}}}}\right)^2,
\end{equation}

where $(\PT^{\text{reco}})_{i}$
are the scalar transverse momenta of the reconstructed objects,
such as jets and photons that pass the above mentioned
identification criteria, the $(\sigma_{\PT})_{i}$ are the expected
resolutions in each object, the $\sigma_{{E}_{x,y}^{miss}}$ are
the resolution of the $\MET\,$ projection along the x-axis and the y-axis,
and the $(\widetilde{p}_{\mathrm T})_{i}$ are the free parameters
allowed to vary in the minimization of the $\chi^2$ function. The
$\widetilde{\mathrm{E}}_{x,y}^{\text{miss}}$ terms are functions of
the free parameters $\widetilde{p}_{x,y}$,
\begin{equation}
  \widetilde{E}_{x,y}^{\text{miss}} =
                           E^{\mathrm{miss,reco}}_{x,y} + \sum_{i=\text{objects}}
                           (p_{x,y}^{\text{reco}})_i-(\widetilde{p}_{x,y})_i .
\end{equation}

In events with no genuine $\MET$, the mismeasured quantities are re-distributed back into the
particle momenta, to minimize the $\chi^{2}$ value. Events are
rejected if the minimized  $\MET$ ($\widetilde{\mathrm{E}}_{\mathrm T}^{\text{miss}}$) is less than 45\GeV and the chi-square
probability is larger than $10^{-3}$.

To further suppress multijet backgrounds, events are not considered if the scalar sum of the transverse
momenta of the identified jets in the event ($\HT$) is greater than 100\GeV. An
additional requirement is applied on the angle ($\alpha$) between the beam direction and the major
axis of the supercluster~\cite{Khachatryan:2015iwa} in order to reject non-prompt photons that have showers
elongated along the beam line.

Finally, the transverse mass, $\MT \equiv \sqrt{\smash[b]{2 \etg \MET [1-\cos\Delta\phi(\gamma,\MET)]}}$,
formed by the photon candidate, $\vmet$, and their opening angle, is required to be greater than
100\GeV. Photons from the continuum $\cPZ\gamma$ background have a harder spectrum than the photons
resulting from the Higgs decay in the SUSY benchmark models considered.
To further reduce the continuum $\cPZ\gamma$ background and for models with
$m_\PH = 125\GeV$ a cut of $\etg < 60\GeV$  is applied. For higher masses the cut is optimized
depending on each mass hypothesis going from 60\GeV up to 200\GeV for $m_\PH = 400\GeV$.

The list of selection criteria used in the model independent and the SUSY benchmark model analyses
are given in Table~\ref{tab:cuts}, together with the cumulative efficiencies relative to the
preselection for signal and background processes.

\begin{table*}[htbp]
\setlength\extrarowheight{2pt}
\centering
\topcaption{Summary of $\Pg\Pg\PH$ selection for both the quasi model-independent analysis and the analysis
with the SUSY benchmark model with the cumulative efficiencies of the selection requirements
relative to the preselection for $\cPZ\Pgg\to\Pgn\Pagn\Pgg$,
$\Pgg$+jet and for a signal in a SUSY benchmark model with $\Pg\Pg\PH$ production of a Higgs boson with
mass 125\GeV decaying into a neutralino of mass 120\GeV and a photon.}
\label{tab:cuts}
\begin{tabular}{lc{c}@{\hspace*{5pt}}c*{3}{c} }
\hline
\multirow{2}{*}{Selection requirements}&  \multicolumn{2}{c}{Model-independent} && \multicolumn{3}{c}{SUSY benchmark model} \\
\cline{2-3}\cline{5-7}
 & $\cPZ\Pgg\to\Pgn\Pagn\Pgg$ & $\Pgg$+jet && $\cPZ\Pgg\to\Pgn\Pagn\Pgg$ & $\Pgg$+jet & $m_{\PSGczDo} = 120\GeV$ \\
\hline
Number of jets $<$2 & 0.909 & 0.769 && \NA & \NA & \NA  \\
$\Delta \phi(\gamma,\text{jet}) < 2.5$ radians & 0.834 & 0.262 && \NA & \NA & \NA  \\
$\text{Transverse mass} > 100$\GeV	& \NA & \NA && 0.867  & 0.292 & 0.829  \\
$\HT <  100$\GeV				& \NA & \NA && 0.785 & 0.188 & 0.804  \\
$\widetilde{E}_{T}^{\text{miss}} > 45$\GeV	& \NA & \NA && 0.761 & 0.071 & 0.743 \\
$\text{Prob}(\chi^2)< 10^{-3}$	& \NA & \NA && 0.626 & 0.033 & 0.467  \\
$\MET\text{ significance}>  20$			& \NA & \NA && 0.440 & 0.001 & 0.195  \\
$\alpha >  1.2$ 				& \NA & \NA && 0.390 & 0.001 & 0.165  \\
$\etg < 60$\GeV		& \NA & \NA && 0.074 & 0.0002 & 0.106  \\\hline
\end{tabular}
\end{table*}

\subsection{Event selection in the \texorpdfstring{$\Z\PH$}{ZH} channel}
The leptonic decays of the $\cPZ$ boson, consisting of two oppositely charged
same-flavor high-$\pt$ isolated leptons ($\Pep \Pem$, $\PGmp \PGmm$), are used
to tag the Higgs boson candidate events. Large missing transverse energy from the undetectable
particles, at least one isolated high-$\et$ photon, and little or moderate jet activity are
required to select the signal events.

The details of the lepton candidate selection and missing transverse energy reconstruction are given in Ref.~\cite{hig13023}.
In addition, photon requirements based on a multivariate selection discussed in Refs.~\cite{Khachatryan:2015iwa,hig13001} have been used. The kinematic
selection requires two leptons with $\pt>20\GeV$ and one photon with $\etg>20\GeV$.
Furthermore, the dilepton mass must be compatible with that of a $\cPZ$ boson within $15\GeV$ of the
pole mass.

To reduce the background from $\PW\cPZ$ events, events are removed if an additional loosely identified
lepton is reconstructed with $\pt > 10\GeV$. To reject most of the top-quark background, an
event is rejected if it passes the b-tagging selection (anti
b-tagging) or if there is a selected jet with $\pt$
larger than $30\GeV$ (jet veto). The b-tagging selection is based on the presence of a muon in the
event from the semileptonic decay of a bottom-quark, and on the impact parameters of the constituent
tracks in jets containing decays of bottom-quarks~\cite{btag}. The set of b-tagging veto criteria retain
about 95\% of the light-quark jets, while rejecting about 70\% of the b-jets.

The signal topology is characterized by a $\cPZ (\ell\ell)$ system with large
transverse momentum balanced in the transverse plane by a $\vmet$ +
$\vg$ system from the Higgs boson decay. To reject background from
$\cPZ\gamma$ and $\cPZ+$jets events with misreconstructed $\MET$ the azimuthal angle
$\Delta \phi_{\ell\ell,\vmet+\vg}$ is required to be greater than 2.7 radians,
the variable $\abs{\pt^{\vmet+\vg}-\pt^{\ell\ell}}/\pt^{\ell\ell}$ is required to be
smaller than 0.5, and the azimuthal angle between the two leptons $\delphill$
is required to be smaller than 2.25 radians. Finally, $\pt^{\ell\ell}$  is
required to be larger than $60\GeV$, and $\MET$ is required to be larger
than $60\GeV$. A summary
of the selection for the analysis is shown in Table~\ref{tab:selectioncuts}.

The signal-to-background fraction depends on the $\abs{\eta^\gamma}$, the pseudorapidity
of the photon, with  greater discrimination
at lower values. To exploit this effect and improve sensitivity, the selected events are
subdivided according to whether the photon is reconstructed in the
barrel or endcap regions, as explained in Section~\ref{sec:results-model-specific}.

\begin{table}[htbp]
  \centering
  \topcaption{Summary of $\cPZ\PH$ selection.}
  \label{tab:selectioncuts}
  \begin{tabular} {lc}
\hline
  Variable & Selection\\
  \hline
Leptons                                           & 2 leptons, $\pt> 20\GeV$ \\
Photons                                           & 1 photon, $\etg> 20\GeV$ \\
$\abs{\mll - m_{\Z}}$                             & $<$15\GeV  \\
Anti b-tagging                                    & applied  \\
Jet veto                                      & 0 jets with $\pt^{\rm j} > 30\GeV$  \\
$\Delta \phi_{\ell\ell,\vmet+\vg}$                &  $>$2.7 radians \\
$\abs{\pt^{\vmet+\vg}-\pt^{\ell\ell}}/\pt^{\ell\ell}$ & $<$0.50  \\
$\delphill$                                       & $<$2.25 radians \\
$\pt^{\ell\ell}$                                   & $>$60\GeV      \\
$\MET$                                           & $>$60\GeV       \\
  \hline
  \end{tabular}
\end{table}

\section{Background estimation} \label{sec:backgrounds}

The background estimation techniques and the composition of all backgrounds in
the search with the $\Pg\Pg\PH$ and $\cPZ\PH$ signatures are discussed below.
The yield for the irreducible background from $\PH\to\cPZ\gamma\to\nu\nu\gamma$
is negligible and is therefore ignored in the analysis.

\subsection{Background estimation in the \texorpdfstring{$\Pg\Pg\PH$}{ggH} channel}

The dominant background for the $\Pgg+$\MET\ signal in the $\Pg\Pg\PH$ channel
is the process $\Pgg$+jet. Other SM backgrounds
include $\cPZ\Pgg\to\Pgn\Pagn\Pgg$, $\PW\Pgg$, $\PW\to \Pe\Pgn$, $\PW\to \mu\Pgn$,
$\PW\to \tau\Pgn$, multijet,
and diphoton events. Background events that do not arise from pp collisions
are also considered in the analysis. These backgrounds can be categorized broadly
into three categories, as described below.

\subsubsection{Background estimates from simulation}

The $\gamma+$jet process surviving the various $\MET$ selection requirements
is one of the most significant backgrounds in this analysis
due to the presence of an isolated photon and its large production cross section.
The MC normalization of this background is corrected using control samples in
data for two event classes, events without jets and those with one or more jets.
The control samples in data are obtained using events collected with a prescaled
single-photon trigger and with the $\MET$ requirement reversed to ensure
orthogonality to the signal phase space, where the contamination from other processes is minimal.
Multiplicative correction factors ($C$) are obtained after
normalizing the event yield in the simulation to match the data in the
control region, separately for events with no jets ($C=1.7$) and one or more
jets ($C=1.1$). These correction factors are used to normalize the simulated
$\gamma+$jet event yield in the signal region. An uncertainty
of 16\% is obtained for these correction factors based on the difference
between the corrected and uncorrected simulation and the relative fraction of no
jet events (about 10\% of the events in the control region) and one or more jet
events. The background processes $\cPZ\Pgg\to\ell\ell\Pgg$ and
$\PW\to \mu\Pgn$ contribute only a small fraction of the total background
prediction, due to the lepton veto applied at the preselection stage,
and are modeled using simulated samples.

To take into account differences between data and simulation due to imperfect
MC modeling, various scale factors (SF) are applied to correct the estimates from simulation.
These SFs are defined by the ratio of the efficiency in data to the
efficiency in simulation for a given selection. The SF for photon reconstruction
and identification is estimated from $\cPZ\to \Pep\Pem$
decays~\cite{CMS:2011aa} and is consistent with unity.

\subsubsection{Background estimates from data} \label{sec:ggh_backgrounds}
The contamination from jets misidentified as photons ($\text{jet} \to \gamma$) is
estimated in a data control sample enriched with multijet events
defined by $\MET < 40\GeV$. This sample is used to measure the ratio
of the number of candidates that pass the photon identification criteria to those failing the isolation requirements.
The numerator of this ratio is further corrected for the photon
contamination due to direct photon production using an isolation side
band. The corrected ratio is applied to data events which
pass the denominator selection and all other event requirements in the
signal region.

The systematic uncertainty of this method is dominated by the choice of the
isolation sideband, and is estimated to be 35\% by changing
the  isolation criteria in the sideband region definition. The other sources
of systematic uncertainty are determined by changing the $\MET\,$ selection
for the control region, and the loose identification
requirements on the photons, all of which are found to be comparatively small.

Events with single electrons misidentified as photons
($\text{electron} \to \gamma$) are another major source
of background. This background is estimated with a tag-and-probe method using
$\cPZ\to \Pep\Pem$ events~\cite{wzxs}. The efficiency to identify electrons
($\epsilon_{\gamma_{\Pe}}$) is estimated in the $\cPZ$ boson peak mass window of
60--120\GeV. The inefficiency ($1-\epsilon_{\gamma_{\Pe}}$)
is found to be $2.31\pm 0.03\%$. The ratio $(1-\epsilon_{\gamma_{\Pe}})/\epsilon_{\gamma_{\Pe}}$,
which represents the electron misidentification rate, is applied to a
sample where candidates are required to have hits in the pixel
detector, and is used to
estimate the contamination in the signal region. The misidentification
rate is found to be dependent on the number of vertices reconstructed in
the event and the number of tracks associated to the selected primary vertex.
The difference in the final yields taking this dependence
into account or neglecting it, using the inclusive measurement of
$\epsilon_{\gamma_{\Pe}}$, is less than 5\%.

\subsubsection{Non-collision background estimates from data}

The search is susceptible to contamination from non-collision backgrounds, which
arise from cosmic ray interactions, spurious signals in the ECAL, and
accelerator-induced secondary particles. The distribution of arrival-times
of photons from these backgrounds is different to that of prompt photons produced in hard scattering. To
quantify the contamination from these backgrounds a fit is performed to the
candidate-time distribution using background distributions from the data~\cite{Khachatryan:2014rwa}.
The contamination due to out-of-time background contributions is found
to be less than one percent of the total background and is therefore not included
in the final event yield.

\subsection{Background estimation in \texorpdfstring{$\cPZ\PH$}{ZH} channel}

Processes that contribute significantly to the SM expectation in the $\cPZ\PH$ channel are listed below.

\subsubsection{Non-resonant dilepton backgrounds}

The contributions from $\PWp\PWm$, top-quark, $\PW+\text{jets}$, and $\Z/\gamma^*\to\tau^+\tau^-$ processes are
estimated by exploiting the lepton flavor symmetry in the final states of
these processes~\cite{CMSinvSearch}. The branching fraction to the
$\Pe^{\pm}\mu^{\mp}$ final state is twice that of the $\Pep\Pem$ or $\PGmp\PGmm$
final states. Therefore, the $\Pe^{\pm}\mu^{\mp}$ control region is used to
extrapolate these backgrounds to the $\Pep\Pem$ and $\PGmp\PGmm$ channels.
The method considers differences between the electron and muon
identification efficiencies. 
The data driven estimates agree well with the number of background events expected when applying the same method to
simulation. The small difference between the prediction and the obtained value
using simulated events is taken as a systematic uncertainty.

The limited number of simulated events is also considered as part of the systematic
uncertainty. In summary, the total systematic uncertainty is about 75\%.
Only two events were selected in the data control region.

\subsubsection{Resonant background with three leptons in the final state}

The $\PW\cPZ \to \ell\nu\ell\ell$ process dominates the resonant
backgrounds with three leptons in the final state. The $\text{electron}\to \gamma$ misidentification rate is
measured in $\Z \to \Pep\Pem$ events by comparing the ratios of electron-electron versus
electron-photon pairs in data and in simulation, as described in Section~\ref{sec:ggh_backgrounds}. The average
misidentification rate is 1--2\% with the larger values at higher $\abs{\eta}^\gamma$.

\subsubsection{Resonant background with two leptons in the final state}

The $\PW\cPZ\to\ell\nu\ell\ell$ process with failure to identify the lepton from
$\PW$ boson decays and the $\cPZ\cPZ \to 2\ell 2\nu$ process dominate
these types of events. The jet $\to \gamma$ misidentification rate is measured in a
sample containing a muon and a photon.
This sample is expected to be dominated by jets misidentified as photons,
with some contamination from $\PW/\cPZ\gamma$ events, which are
subtracted in the study using the simulated prediction. The
misidentification rate is similar to the obtained values in the $\Pg\Pg\PH$ channel.

\subsubsection{Resonant background with no genuine missing transverse energy}

The background from $\cPZ\gamma$ or $\cPZ+\text{jets}$ events
is predicted by the simulation to be about 15\% of the total background.
Several data regions are studied to verify that the background is
estimated correctly. A good agreement between data and simulation is
found in all cases. A 50\% uncertainty, the statistical uncertainty of the control region, is
taken for these backgrounds estimated from simulation.

\section{Summary of systematic uncertainties} \label{sec:systematics}
Systematic uncertainties in the background estimates from control samples
in data are described in Section~\ref{sec:backgrounds}.
A summary of the systematic uncertainties considered in each channel
are listed in Tables~\ref{tab:sys} and~\ref{tab:syst}.

A common source of systematic uncertainty is associated with the
measurement of the integrated luminosity, determined to
2.6\%~\cite{CMS:2013gfa}.
The uncertainties in the normalization of signal and simulation-based  backgrounds are obtained
by varying the renormalization and factorization scales, and the parton distribution
functions~\cite{Botje:2011sn,Alekhin:2011sk,Lai:2010vv,Martin:2009iq,Ball:2011mu,MCFM}.

Because the model-independent and model-specific selections differ
significantly in the $\Pg\Pg\PH$ channel, the systematic uncertainties are
evaluated separately for each selection.
The photon energy scale uncertainty~\cite{Khachatryan:2015iwa} of about 1$\%$ affects the
signal and background predictions by 4\% for the model specific
selection and by 0.5\% for the model-independent selection.
Similarly, the jet energy scale uncertainty affects the signal and
background predictions by 2--5\% depending on the process and
selection.  After changing the photon or jet energy scales, the $\MET$
is also recomputed. In addition, the systematic uncertainty associated
with the jet energy resolution and unclustered energy scale
are propagated to the $\MET\,$ computation, and affect the signal and
background predictions by 2--4\%. As
described in the previous section, a 16\% uncertainty is applied to
the $\gamma$+jet normalization due to the difference in the jet
multiplicity distribution between the data and background predictions
in the  $\gamma$+jet control region. The uncertainty due to the pileup
modeling is found to be 1\%, and is estimated by shifting the central
value of the total inelastic cross section within its
uncertainty.

\begin{table*}[htbp]
\centering
\topcaption{Summary of all relative systematic uncertainties
  in percent for the signal and background estimates for the
  Higgs model (model-independent in parenthesis) selection in the $\Pg\Pg\PH$ analysis.}
\label{tab:sys}
{
\begin{tabular}{lcccccc}
\hline
Source   & Signal & Jet$\to \gamma$ & Electron$\to \gamma$ & $\gamma$ + jet & $\cPZ\nu\nu\gamma$ & $\PW\gamma$ \\
\hline
PDF & 10 (0) & \NA & \NA & \NA & 4 (4) & 4 (4) \\
Integrated luminosity & 2.6 (2.6) & \NA & \NA & 2.6 (2.6) & 2.6 (2.6) & 2.6 (2.6) \\
Photon efficiency& 3 (3) & \NA & \NA & 3 (3) & 3 (3) & 3 (3) \\
Photon energy scale $\pm$ 1 $\%$  & 4 (0.5) & \NA & \NA & 4 (0.5) & 4 (0.5) & 4 (0.5) \\
$\MET$ energy scale & 4 (2) & \NA & \NA & 4 (2) & 4 (2) & 4 (2) \\
Jet energy scale & 3 (2) & \NA & \NA & 5 (5) & 3 (2) & 3 (2) \\
Pileup  &  1 (1) & \NA & \NA & 1 (1) & 1 (1) & 1 (1) \\
$\cPZ\nu\nu\gamma$ normalization & \NA & \NA & \NA & \NA & 3 (3) & \NA \\
$\gamma + \text{jet}$ normalization  & \NA & \NA & \NA & 16 (16) & \NA & \NA \\
$\PW\gamma$ normalization & \NA & \NA & \NA & \NA & \NA & 3 (3) \\
$\text{Jet}\to \gamma$    & \NA & 35 (35) & \NA & \NA & \NA & \NA \\
$\text{Electron}\to \gamma$ & \NA & \NA & 6 (6) & \NA & \NA & \NA \\
\hline
\end{tabular}
}
\end{table*}

In the $\cPZ\PH$ channel,
lepton-reconstruction and identification scale factors are measured
using a control sample of $\dyll$ events in the $\cPZ$ peak region~\cite{wzxs}.
The associated uncertainty is about 2\% per lepton. The photon identification
uncertainty is taken to be 3\%~\cite{hig13001}.
The effect of uncertainties in jet-energy scale and $\MET$ on the analysis is
also considered. The uncertainty in the b-tagging efficiency is
estimated to be about 0.7\% comparing inclusive $\dyll$ samples in data
and simulation.
The total uncertainty in the
background estimates in the signal region is 36\%, which is
dominated by the statistical uncertainties in the data control samples from which they are derived.

The impact of the systematic uncertainties in the $\Pg\Pg\PH$ channel is
relatively important: the sensitivity is increased by about 50\%
if all the systematic uncertainties are removed, where the
normalization uncertainties on the $\gamma + \text{jet}$ and
$\text{jet}\to \gamma$ background processes dominate.
The $\cPZ\PH$ channel is limited by the statistical uncertainty, and the effect of the systematic uncertainties reduce the sensitivity by less than 10\%.

\begin{table*}[htbp]
  \centering
\topcaption{\label{tab:syst} Summary of relative systematic uncertainties
in percent for the signal and background estimates in the $\cPZ\PH$ analysis.}
{
  \begin{tabular}{lccccc}
\hline
 Source & $\cPZ\PH$ & $\cPZ\gamma$ or $\Zjets$ & $\PW\cPZ$ & $\cPZ\cPZ$ & $\PW\PW$ + top-quark \\
\hline
Integrated luminosity                           &   2.6  &   \NA    &  2.6    &  2.6  &	 \NA    \\
Lepton efficiency                               &   3.6  &   \NA    &  3.6    &  3.6  &	 \NA    \\
Photon efficiency                               &   3.0  &   \NA    &   \NA     &   \NA   &    \NA    \\
Momentum resolution                             &   0.5  &   \NA    &  1      &  1    &	 \NA    \\
$\MET$ energy scale                             &   0.5  &   \NA    &  0.6    &  0.1  &	 \NA    \\
Jet energy scale                                &   2	 &   \NA    &  4      &  4    &	 \NA    \\
b-tagging                                       &   0.7  &   \NA    &  0.7    &  0.7  &	 \NA    \\
Underlying event                                &   3	 &   \NA    &   \NA     &   \NA   &    \NA    \\
PDF                                             &   7.1  &   \NA    &  6.3    &  7.7  &	 \NA    \\
Renorm. and factor. scales                      &   7.0	 &   \NA	  & 10.7    &  6.5  &    \NA    \\
 $\dyll$ normalization                          &   \NA	 &  50    &    \NA    &	 \NA  &	 \NA    \\
Non-resonant dilepton bkg. norm. &   \NA    &   \NA    &    \NA    &    \NA  &  70     \\
$\text{Jet}\to \gamma$                          &   \NA    &   \NA    & 30      &  30   &    \NA    \\
$\text{Electron}\to \gamma$                     &   \NA    &   \NA    & 10      &  10   &    \NA    \\
Amount of simulated events                      &   3.5  &  60    & 10      &  30   &  40     \\
\hline
  \end{tabular}
}

\end{table*}

Correlations between systematic uncertainties in the two channels are taken
into account. In particular, the main sources of correlated systematic
uncertainties are those in the experimental measurements such as the integrated
luminosity, photon identification, the jet energy scale, and missing transverse
energy resolution. All other systematic uncertainties are uncorrelated between
them given the different signal and background processes.

\section{Results} \label{sec:results}

The results from the two searches and their combination are
reported in this section. In the absence of deviations from the standard
model predictions, the modified frequentist method,
$\mathrm{CL_s}$~\cite{LHC-HCG-Report,Read1,junkcls}, is used to define the exclusion limits.

\subsection{Model-independent results in the \texorpdfstring{$\Pg\Pg\PH$}{ggH} channel}

\begin{figure}[htbp]
\centering
\includegraphics[width=0.49\textwidth]{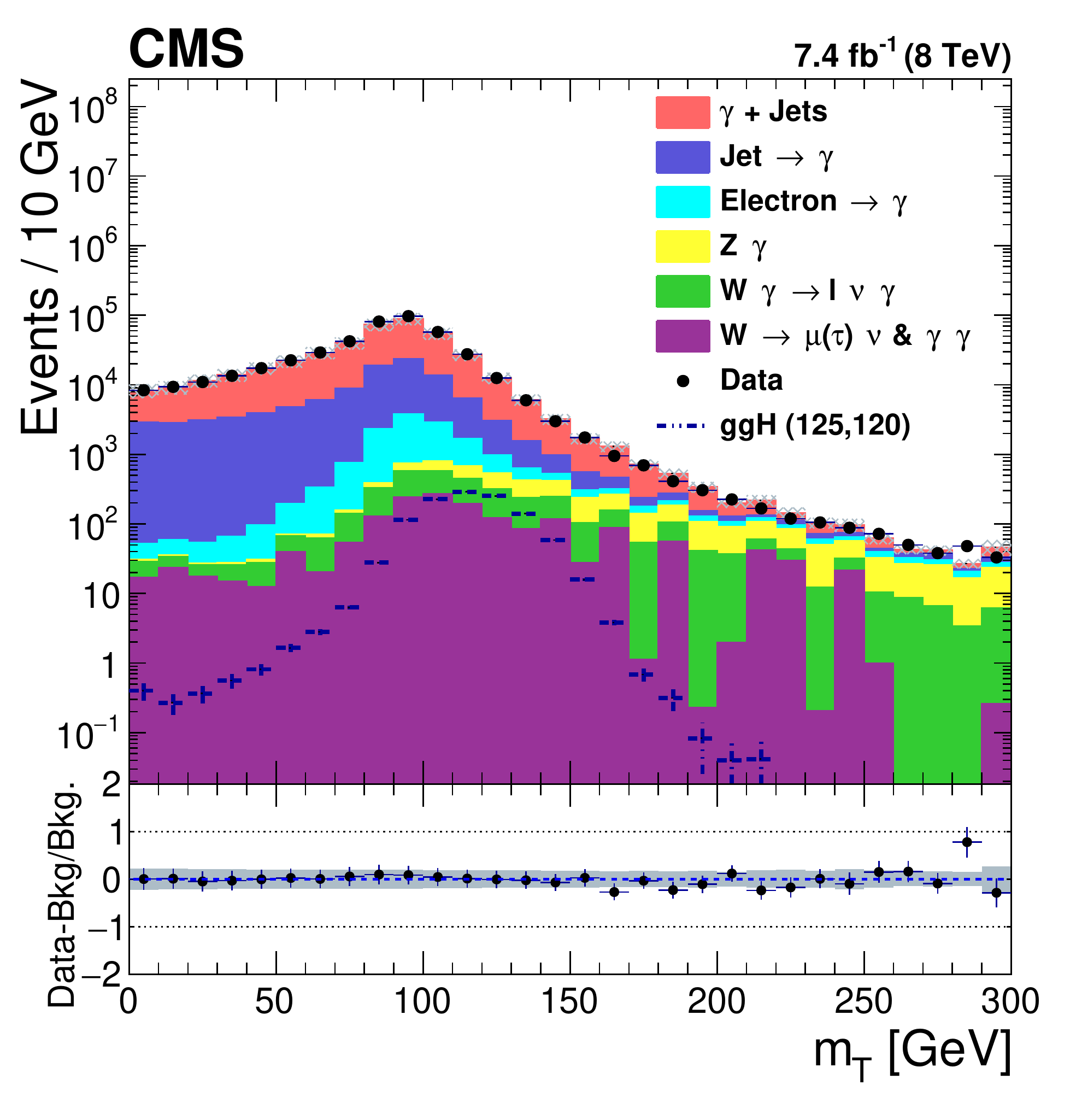}
\includegraphics[width=0.49\textwidth]{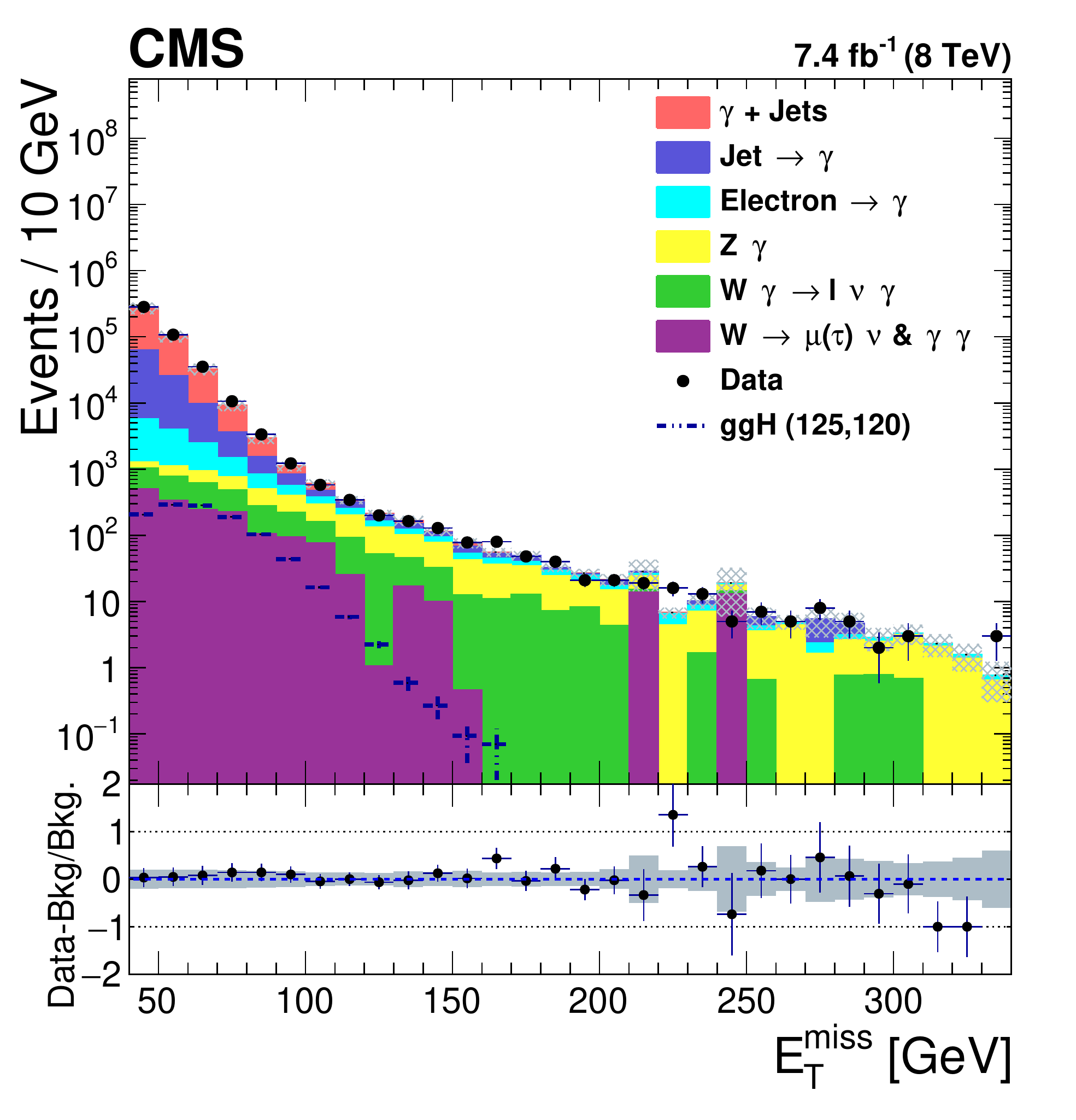}
\caption{The $\MT$ and $\MET$ distributions for data, background
  estimates, and signal after the model-independent selection for the
  $\Pg\Pg\PH$ channel.
The bottom panels in each plot show the ratio of
  ($\text{data}$ - $\text{background}$)/background and the gray band includes both the
  statistical and systematic uncertainties in the background
  prediction.
The signal is shown for $\mH = 125\GeV$ and $m_{\PSGczDo} =  120\GeV$ and a 100\% branching fraction.}
\label{fig:modelInd}
\end{figure}

Because of the variety of possible BSM signals that could contribute to this final
state, the results are presented for a signal with the model-independent selection
described in Section~\ref{sec:selection}. The total number of observed
events and the estimated SM backgrounds are summarized in
Table~\ref{table:modelInd}, and found to be compatible within their
uncertainties.
Figure~\ref{fig:modelInd} shows the $\MT$ and $\MET$ distributions for the
model-independent selection.

\begin{table}[hbtp]
\center
\topcaption{Observed yields and background estimates at
 $8\TeV$ in the $\Pg\Pg\PH$ channel after the model-independent selection.
 Statistical and systematic uncertainties are shown.}
\label{table:modelInd}
{
\begin{tabular}{lc}
\hline
Process & Event yields \\
\hline
$\gamma +$ jets                          & $(313 \pm 50 ) \times 10^3$ \\
$\text{jet}\to \gamma$        & $(910 \pm 320 ) \times 10^2$ \\
$\Pe \to \gamma$         & $10350 \pm 620$ \\
$\PW(\to \ell\nu)+\gamma $                 &  $2239 \pm 111$ \\
$\cPZ( \to \nu \bar{\nu} )+\gamma    $      &  $2050 \pm 102$ \\
Other                                    &  $1809 \pm 91$ \\
\hline
Total background                       &   $(420 \pm 82 ) \times 10^3$ \\
\hline
Data                                   &  $442 \times 10^3$  \\
\hline
\end{tabular}
}
\end{table}

\begin{figure}[htb]
\centering
\includegraphics[width=\cmsFigWidth]{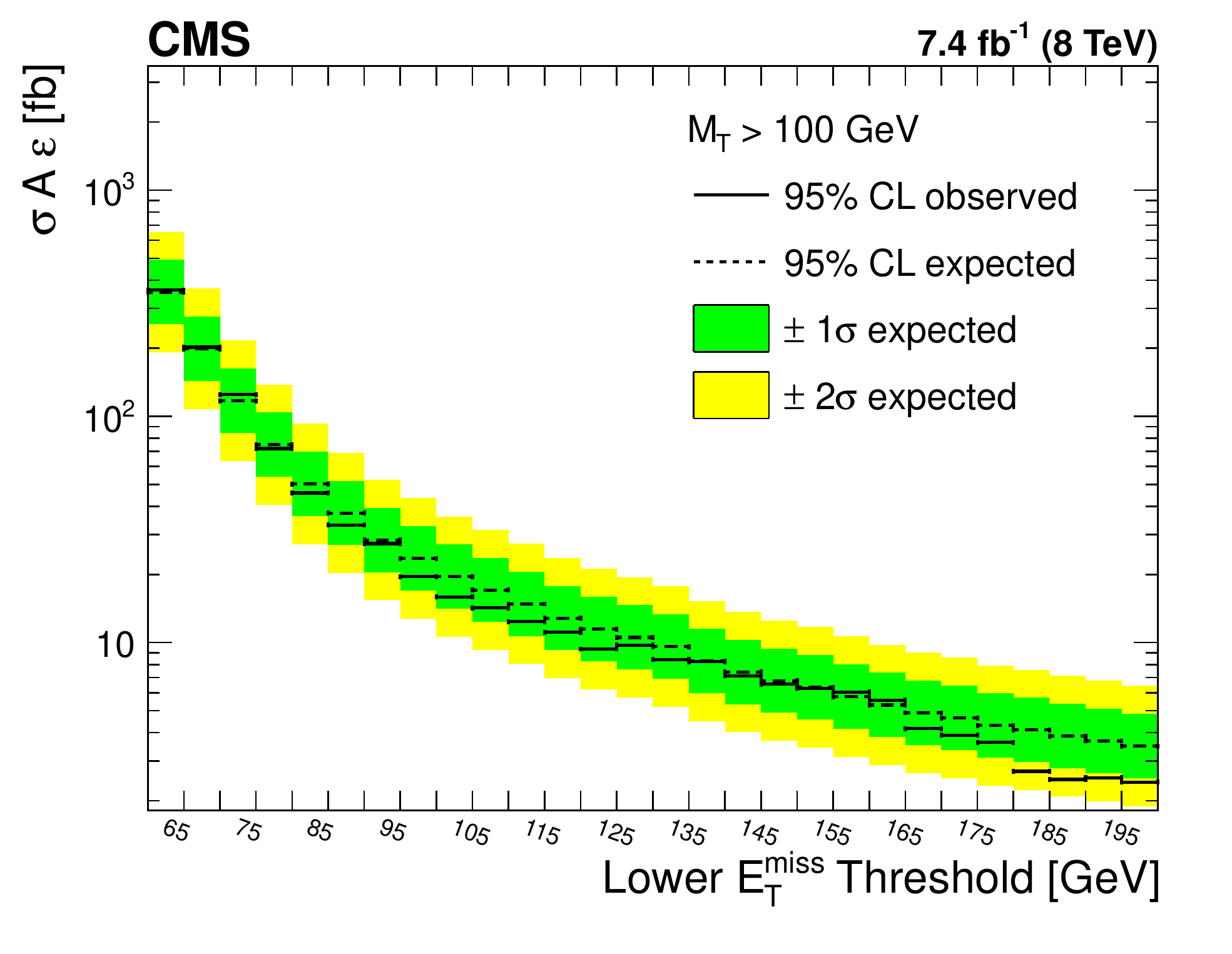}
\caption{ The expected and observed 95\% CL upper limit on the product of cross
section, acceptance, and efficiency ($\sigma (\Pp\Pp
\to \gamma + \MET) A \epsilon$) for $\MT> 100\GeV$, as function of
the $\MET$ threshold for the $\Pg\Pg\PH$ channel.}
\label{fig:limit_MI}
\end{figure}

Figure~\ref{fig:limit_MI} shows the observed and expected
model-independent 95\% CL upper limits for the $\Pg\Pg\PH$ analysis
on the product of cross section, acceptance, and
efficiency for $\MT>100\GeV$, as a function of $\MET$ threshold.

\subsection{Model-specific results in the \texorpdfstring{$\Pg\Pg\PH$}{ggH} channel}\label{sec:results-model-specific}
\begin{figure}[htbp]
\centering
\includegraphics[width=0.49\textwidth]{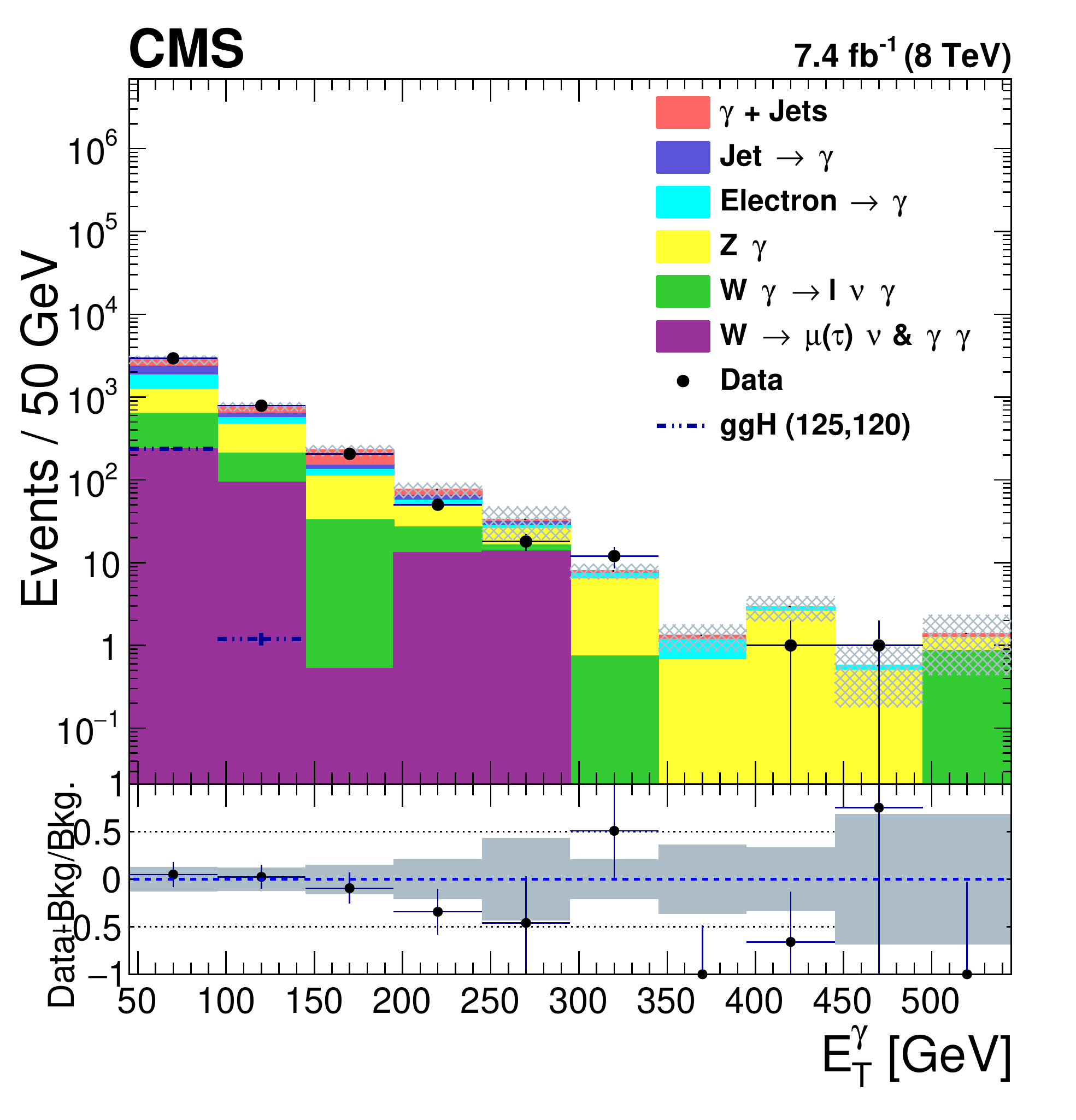}
\caption{The transverse energy distribution of photons for data, the background estimates, and signal after the model-dependent selection (except the upper  selection on the photon) for the $\Pg\Pg\PH$ channel. The bottom panel shows the ratio of ($\text{data}$ - $\text{background}$)/background and the gray band includes both the statistical and systematic uncertainties in the background prediction. The signal is shown for $\mH = 125\GeV$ and $m_{\PSGczDo} = 120\GeV$.
    \label{fig:modeli}}
\end{figure}

Imposing the model-specific selection described in Section~\ref{sec:selection}
for the $\Pg\Pg\PH$ channel, 1296 events are selected in data with a total estimated background of $1232\pm188$. The yields for this selection
are shown in Table~\ref{tab:exoh} and estimated for Higgs boson decays ($\PH\to \PXXSG\PSGczDo,\PSGczDo\to
\PXXSG\gamma$) assuming the $\Pg\Pg\PH$ production rate for SM Higgs bosons and
a 100\% branching fraction for this decay. Figure~\ref{fig:modeli}
shows the transverse energy
distribution of photons for data, the background estimates, and signal
after the model-dependent selection, except the upper selection on the
photon, for the $\Pg\Pg\PH$ channel.

\begin{table*}[htbp]
 \centering
 \topcaption{Observed yields, background estimates, and signal predictions at
 $8\TeV$ in the $\Pg\Pg\PH$ channel for different values of the $m_{\PSGczDo}$ and for
  different $c\tau_{\PSGczDo}$ of the $\PSGczDo$. These correspond to
 $\mathcal{B}(\PH \to \text{undetectable}+\gamma)=$100\%, assuming the SM
 cross section at the given $\mH$ hypothesis.
 The combination of statistical and systematic uncertainties is shown for the yields.}
\label{tab:exoh}
 \begin{tabular}{lc}
\hline
Process & Event yields \\
\hline
$\Pg\Pg\PH(\mH = 125\GeV,m_{\PSGczDo} =  65\GeV)$ &   653 $\pm$  77 \\
$\Pg\Pg\PH(\mH = 125\GeV,m_{\PSGczDo} =  95\GeV)$ &  1158 $\pm$ 137 \\
$\Pg\Pg\PH(\mH = 125\GeV,m_{\PSGczDo} = 120\GeV)$ &  2935 $\pm$ 349 \\
\hline
$\Pg\Pg\PH(\mH = 125\GeV,m_{\PSGczDo} =  95\GeV)$ $c\tau_{\PSGczDo} = 100$\unit{mm}   &  983 $\pm$ 116  \\
$\Pg\Pg\PH(\mH = 125\GeV,m_{\PSGczDo} =  95\GeV)$ $c\tau_{\PSGczDo} = 1000$\unit{mm}  &  463 $\pm$  55  \\
$\Pg\Pg\PH(\mH = 125\GeV,m_{\PSGczDo} =  95\GeV)$ $c\tau_{\PSGczDo} = 10000$\unit{mm} &   83 $\pm$  10  \\
\hline
$\Pg\Pg\PH(\mH = 150\GeV,m_{\PSGczDo} = 120\GeV)$ &  4160 $\pm$ 491  \\
$\Pg\Pg\PH(\mH = 200\GeV,m_{\PSGczDo} = 170\GeV)$ &  5963 $\pm$ 704  \\
$\Pg\Pg\PH(\mH = 300\GeV,m_{\PSGczDo} = 270\GeV)$ &  5152 $\pm$ 608  \\
$\Pg\Pg\PH(\mH = 400\GeV,m_{\PSGczDo} = 370\GeV)$ &  4057 $\pm$ 479  \\
\hline
$\gamma +$ jets                          &  179 $\pm$  28 \\
$\text{jet}\to \gamma$            &  269 $\pm$  94 \\
$\Pe \to \gamma$             &  355 $\pm$  28 \\
$\PW(\to \ell\nu)+\gamma $                &  154 $\pm$  15 \\
$\cPZ( \to \nu \bar{\nu} )+\gamma$         &  182 $\pm$  13 \\
Other                                    &   91 $\pm$  10 \\
\hline
Total background                         & 1232 $\pm$ 188 \\
\hline
Data                                     & 1296  \\
\hline
 \end{tabular}

\end{table*}

\subsection{Results in the \texorpdfstring{$\cPZ\PH$}{ZH} channel and combinations}\label{sec:results_zh_combination}

\begin{figure}[htbp]
\centering
\includegraphics[width=0.49\textwidth]{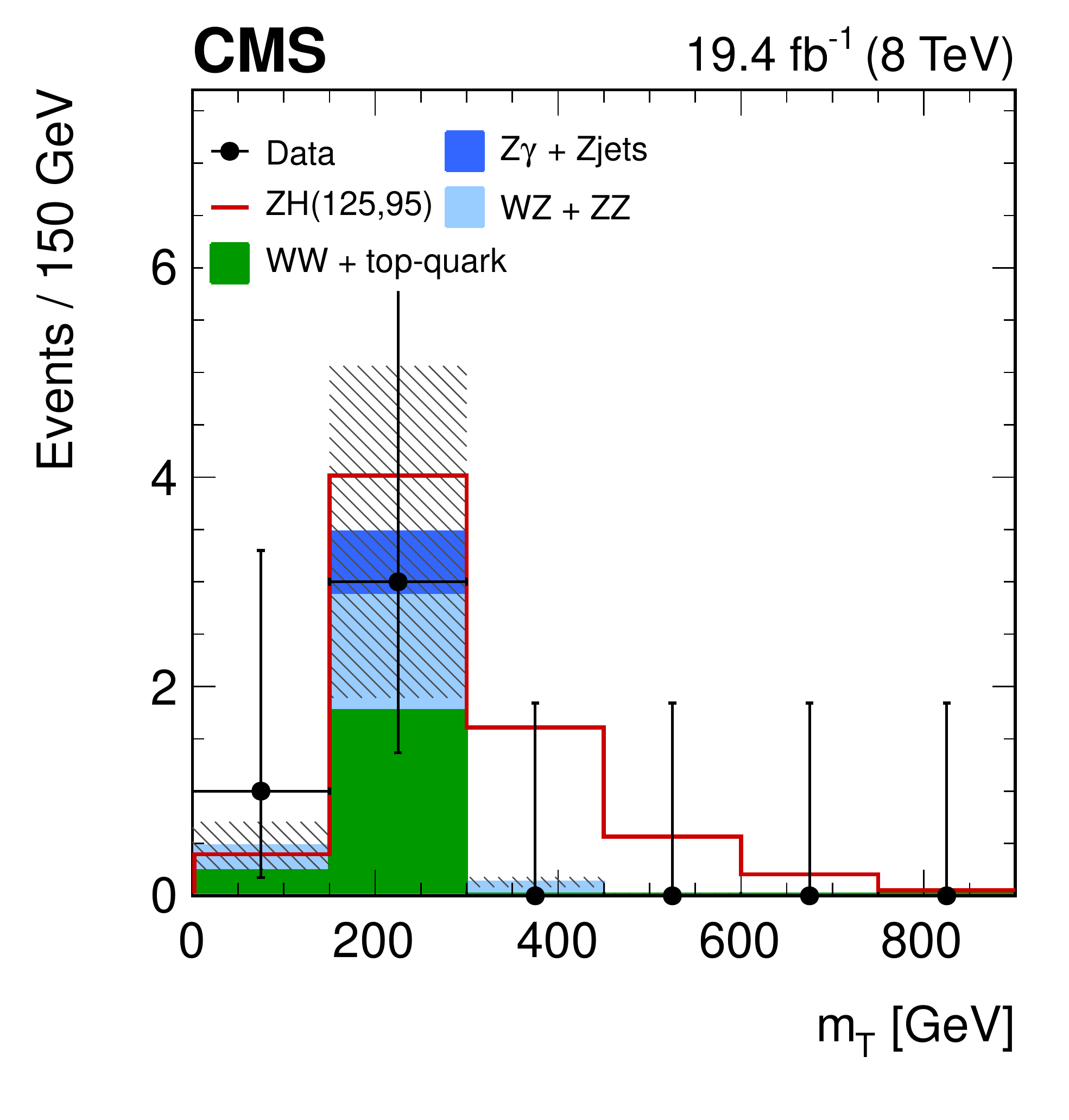}
\includegraphics[width=0.49\textwidth]{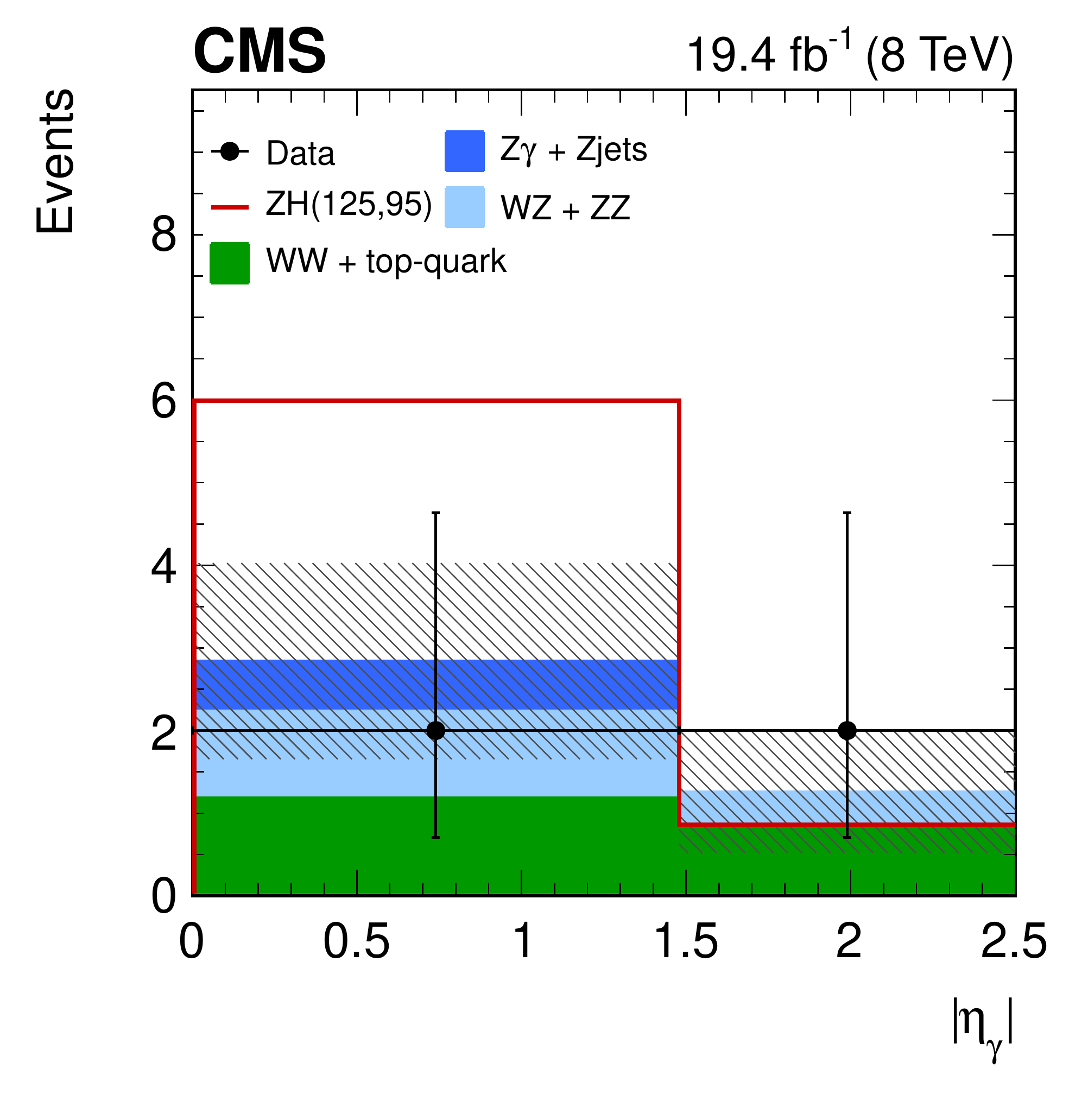}
\caption{Distributions in signal where $\mH = 125\GeV$ and $m_{\PSGczDo} = 95\GeV$,
backgrounds and data for $\mT$ (\cmsLeft)
and $\abs{\eta^\gamma}$ (\cmsRight) after applying all requirements. The uncertainty
band for the backgrounds includes both statistical and systematic uncertainties.
The signal model assumes a SM $\cPZ\PH$ production rate for a Higgs boson with
$\mH = 125\GeV$ and a 10\% branching fraction.
    \label{fig:presel}}
\end{figure}

A total of four events are selected with the search in $\cPZ\PH$. The background yield is estimated to $4.1\pm1.8$.
The numbers of observed and expected events are shown in Table~\ref{tab:zhinvsel}.
The signal model assumes a SM $\cPZ\PH$ production rate and
a 100\% branching fraction to undetectable particles and one
or two photons. The expected signal yield is larger for cases where
$m_{\PSGczDo}$ is smaller than $\mH/2$ since there are two photons in the
final state ($\PH \to \PSGczDo\PSGczDo \to \gamma \gamma \PXXSG\PXXSG$),
and as a result the sensitivity improves for smaller masses.
Good agreement between the data and the background prediction is
observed. The transverse mass,
$\mT \equiv \sqrt{\smash[b]{2\pt^{\ell\ell}\pt^{\vmet+\vg}[1-\cos(\Delta
  \phi_{\ell\ell,\vmet+\vg})]}}$, and $\abs{\eta^\gamma}$ distributions discriminate
  signal and background and
are shown in Figure~\ref{fig:presel} at the final step of the selection.

\begin{table*}[htbp]
 \centering
 \topcaption{Observed yields, background estimates, and signal predictions at
 8 $\TeV$ in the $\cPZ\PH$ channel for different values of the $m_{\PSGczDo}$ and for
  different $c\tau_{\PSGczDo}$ of the $\PSGczDo$. The signal predictions correspond to
 $\mathcal{B}(\PH \to {\rm undetectable}+\gamma)=$100\% assuming the SM $\cPZ\PH$
 cross section at the given $\mH$ hypothesis. The combination of statistical and systematic uncertainties is shown for the yields.}
\label{tab:zhinvsel}
 \begin{tabular}{lc}
\hline
Process & Event yields \\
\hline
$\cPZ\PH(\mH = 125\GeV,m_{\PSGczDo} =   1\GeV)$ &  69.2 $\pm$ 8.4  \\
$\cPZ\PH(\mH = 125\GeV,m_{\PSGczDo} =  10\GeV)$ &  68.6 $\pm$ 8.4  \\
$\cPZ\PH(\mH = 125\GeV,m_{\PSGczDo} =  30\GeV)$ &  53.5 $\pm$ 6.5  \\
$\cPZ\PH(\mH = 125\GeV,m_{\PSGczDo} =  60\GeV)$ &  47.7 $\pm$ 5.8  \\
$\cPZ\PH(\mH = 125\GeV,m_{\PSGczDo} =  65\GeV)$ &  40.0 $\pm$ 4.9  \\
$\cPZ\PH(\mH = 125\GeV,m_{\PSGczDo} =  95\GeV)$ &  40.3 $\pm$ 4.9  \\
$\cPZ\PH(\mH = 125\GeV,m_{\PSGczDo} = 120\GeV)$ &  39.0 $\pm$ 4.8  \\
\hline
$\cPZ\PH(\mH = 125\GeV,m_{\PSGczDo} =  95\GeV)$ $c\tau_{\PSGczDo} = 100$\unit{mm}   &  39.3 $\pm$ 4.8  \\
$\cPZ\PH(\mH = 125\GeV,m_{\PSGczDo} =  95\GeV)$ $c\tau_{\PSGczDo} = 1000$\unit{mm}  &  17.6 $\pm$ 2.2  \\
$\cPZ\PH(\mH = 125\GeV,m_{\PSGczDo} =  95\GeV)$ $c\tau_{\PSGczDo} = 10000$\unit{mm} &   2.6 $\pm$ 0.3  \\
\hline
$\cPZ\PH(\mH = 200\GeV,m_{\PSGczDo} = 170\GeV)$ &  13.1 $\pm$ 1.6  \\
$\cPZ\PH(\mH = 300\GeV,m_{\PSGczDo} = 270\GeV)$ &   3.5 $\pm$ 0.4  \\
$\cPZ\PH(\mH = 400\GeV,m_{\PSGczDo} = 370\GeV)$ &   1.2 $\pm$ 0.1  \\
\hline
$\cPZ\gamma + \Zjets$    &   0.6 $\pm$  0.4  \\
$\PW\cPZ$                 &   1.2 $\pm$  0.3  \\
$\cPZ\cPZ$                 &   0.3 $\pm$  0.1  \\
$\PW\PW$ + top-quark     &   2.0 $\pm$  1.7  \\
\hline
Total background       &   4.1 $\pm$  1.8  \\
\hline
Data                   &  4   	          \\
\hline
 \end{tabular}

\end{table*}

\begin{figure}[htb]
  \centering
   \includegraphics[width=\cmsFigWidth]{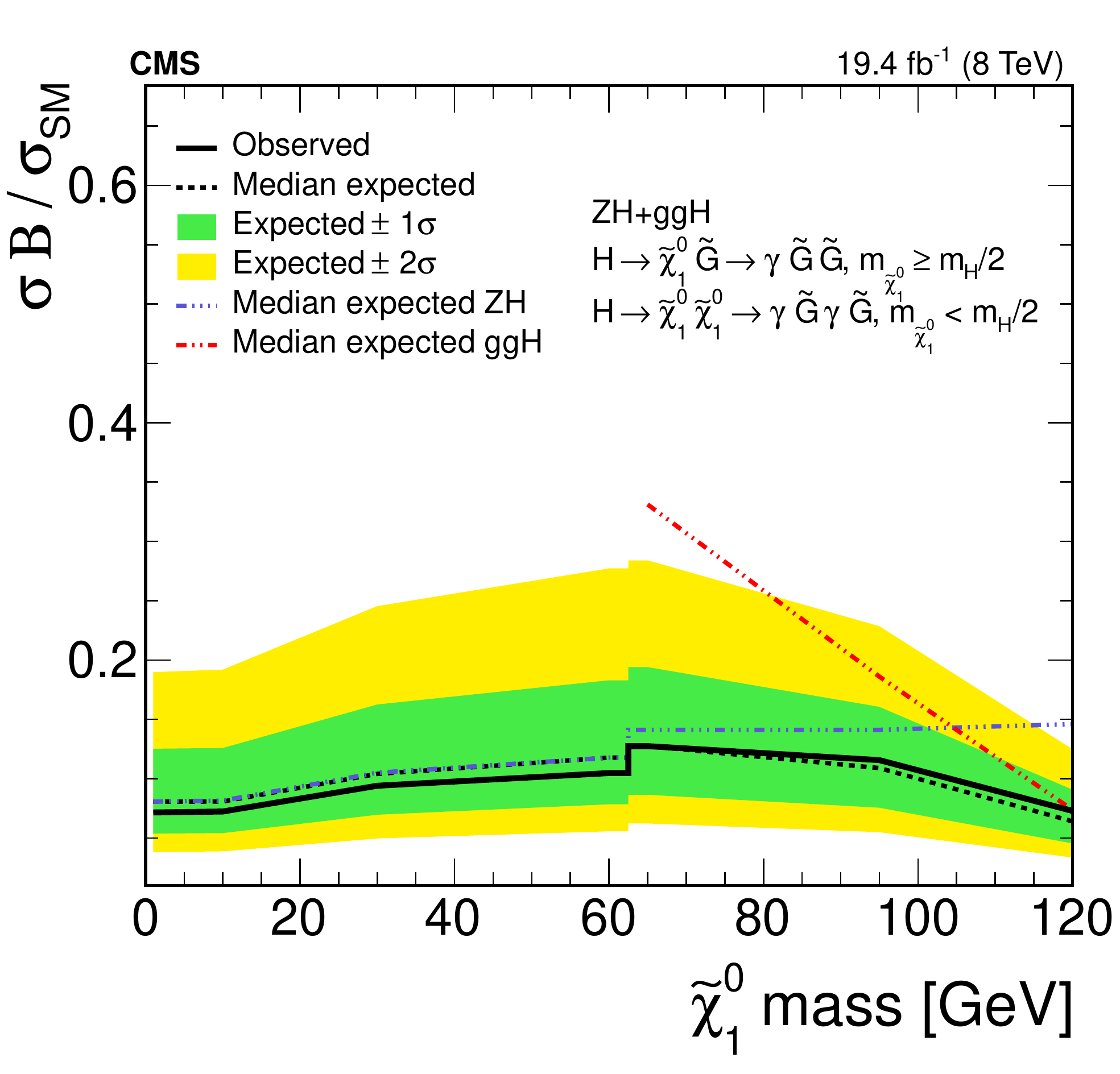}
   \caption{
     Expected and observed 95\% CL upper limits on $\sigma \, \mathcal{B}/\sigma_{ \mathrm{SM}}$ for
     $\mH = 125\GeV$ as a function of $m_{\PSGczDo}$ assuming the SM Higgs boson cross sections,
     for the $\cPZ\PH$ and $\Pg\Pg\PH$ channels and their combination, with $\mathcal{B} \equiv
     \mathcal{B}(\PH\to \PSGczDo\PSGczDo) \,\mathcal{B}(\PSGczDo \to
     \PXXSG+\gamma)^2$ for $m_{\PSGczDo} < \mH/2$ and $\mathcal{B} \equiv \mathcal{B}(\PH\to
     \PSGczDo\PXXSG) \, \mathcal{B}(\PSGczDo \to \PXXSG+\gamma)$ for $m_{\PSGczDo} \geq
     \mH/2$.}
    \label{fig:chi0_mass}

\end{figure}

The 95\% CL upper limits are extracted from counting experiments in three
categories: the model-specific selection in the $\Pg\Pg\PH$ channel, and
photons identified in the barrel and the endcap calorimeters for the $\cPZ\PH$ channel.
Results are combined using a binned-likelihood method.
The 95\% CL upper limits on ($\sigma \, \mathcal{B} )/\sigma_{SM}$, where $\sigma_{SM}$ is the cross
section for the SM Higgs boson, are evaluated for different mass values of $\PSGczDo$ ranging from
$1\GeV$ to $120\GeV$ for the individual searches and their combination and are shown in
Figure~\ref{fig:chi0_mass}. The upper limits for $m_{\PSGczDo} < \mH/2$ are not shown for the
$\Pg\Pg\PH$ channel because the sensitivity is very low due to the combination kinematic properties
and the corresponding selection; in particular the $\MET$ and photon $\pt$ values tend to be outside the selected
ranges. A 95\% CL upper limit on the branching fraction of 10\% is set for a
neutralino mass of $95\GeV$.

\begin{figure}[htbp]
  \centering
\includegraphics[width=0.49\textwidth]{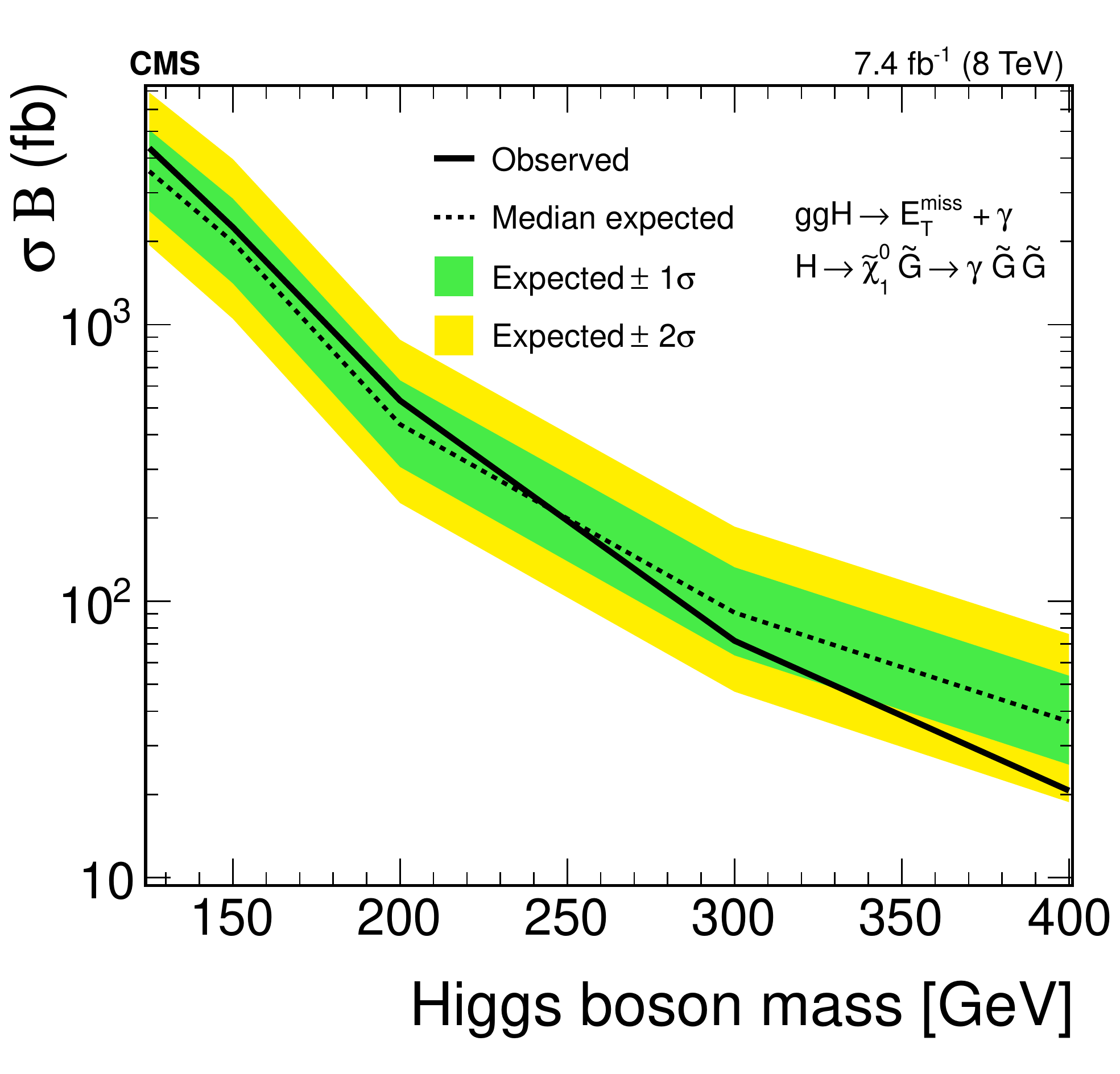}
\includegraphics[width=0.49\textwidth]{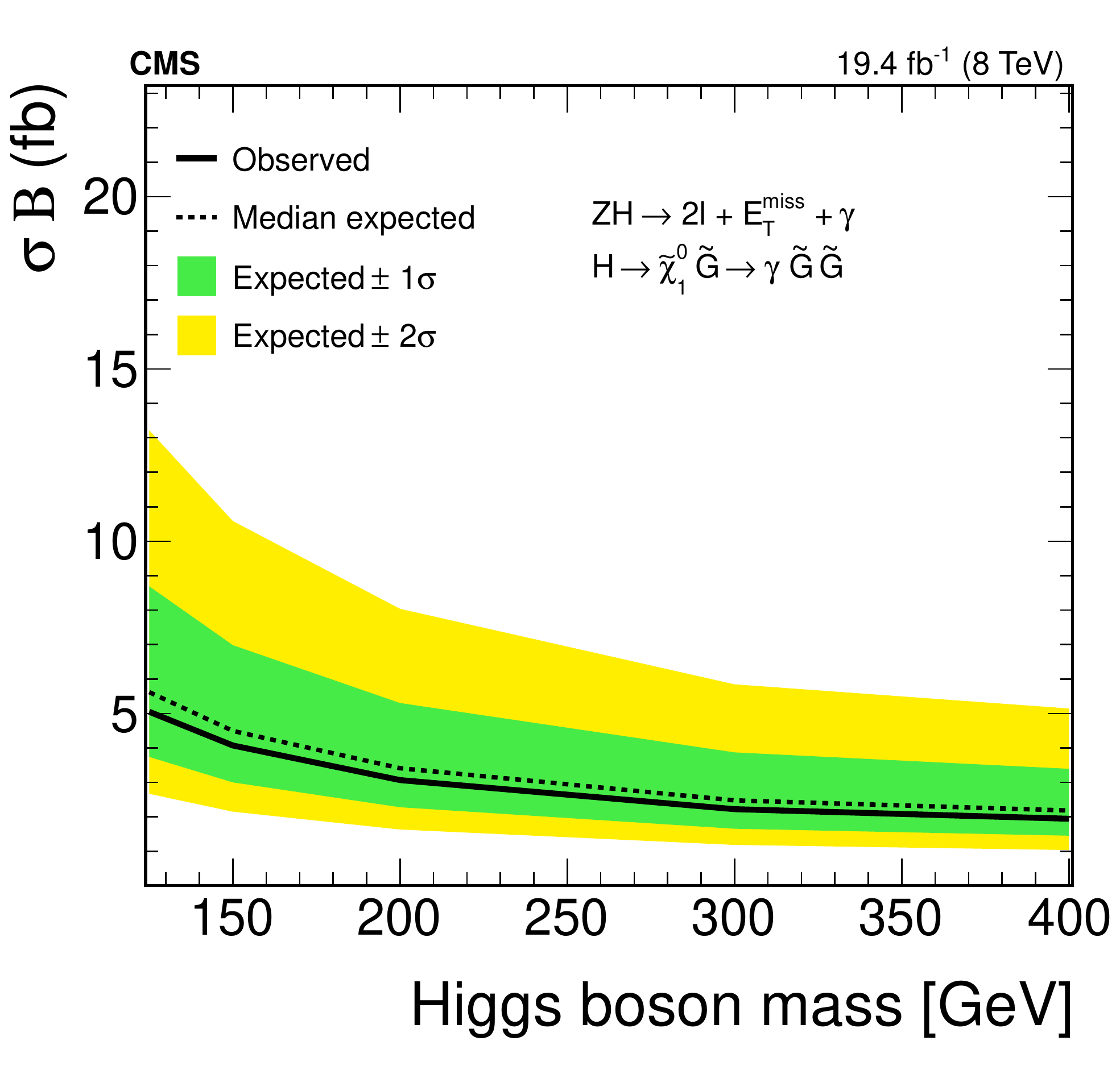}
    \caption{Expected and observed 95\% CL upper limits on $\sigma_{gg
        \to \PH} \, \mathcal{B}$ as a function of the Higgs
      boson mass with $m_{\PSGczDo} = \mH - 30\GeV$ in
      $\Pg\Pg\PH$ channel (\cmsLeft) and in the $\cPZ\PH$ channel (\cmsRight).
    }
    \label{fig:higgs_mass}

\end{figure}

Expected and observed limits are also shown for the decay of possible
heavier scalar Higgs bosons as a function of the Higgs boson mass in Figure~\ref{fig:higgs_mass}.
The requirement on $\etg$ used in the $\Pg\Pg\PH$ channel is
removed. A lower threshold on $\etg$ is added, optimized to maximize the sensitivity for each mass hyposthesis.  A combination of the two channels
is not performed because the assumption of a common SM Higgs boson cross
section is not justified.

\begin{figure}[htbp]
  \centering
\includegraphics[width=\cmsFigWidth]{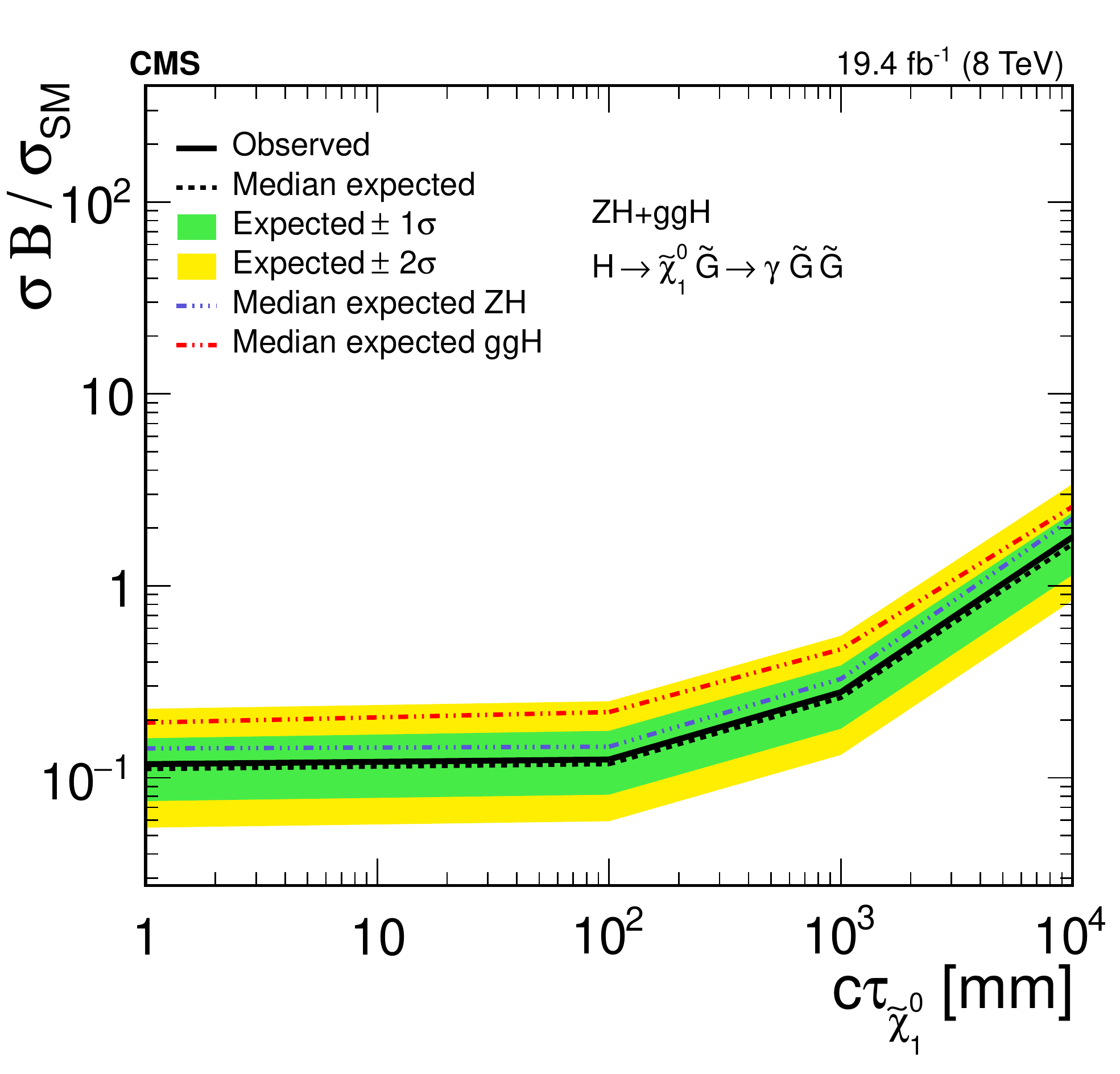}
    \caption{Expected and observed 95\% CL upper limits on $\sigma_{\PH} \,\mathcal{B}$
    as a function of $c\tau_{\PSGczDo}$ for $\mH = 125\GeV$ and $m_{\PSGczDo}=95\GeV$, where
    $\mathcal{B} \equiv \mathcal{B}(\PH\to \PSGczDo\PXXSG)
    \, \mathcal{B}(\PSGczDo \to \PXXSG+\gamma)$.
    }
    \label{fig:chi0_ctau}

\end{figure}

As discussed in the introduction, some BSM models predict $\PSGczDo$
neutralinos with sizable lifetimes. The performance of the searches has been evaluated
for finite lifetimes without modifying the analysis strategy. The
expected and observed limits are shown in Figure ~\ref{fig:chi0_ctau} as function
of $c\tau_{\PSGczDo}$. The results are shown for $\mH = 125\GeV$ and
$m_{\PSGczDo}=95\GeV$.
As seen in Tables~\ref{tab:exoh} and~\ref{tab:zhinvsel}, the selection efficiency is roughly
constant for values of $c\tau_{\PSGczDo}$  less than 10\cm, and drops
rapidly for larger values. The default timing criteria applied in the ECAL energy
reconstruction are the cause for the decrease in the efficiency. In particular,
there is a requirement of a maximum of 3\unit{ns} on the photon arrival time relative to the
nominal time-of-flight for prompt photons. The delayed arrival time of
the photon can be caused by a kink in the trajectory or by a lower
velocity of the neutralino.

\section{Summary}
A search is presented for exotic decays of a Higgs boson into undetectable particles
and one or two isolated photons in $\Pp\Pp$ collisions at a center-of-mass energy of
$8\TeV$. The data correspond to an integrated luminosity of up to
19.4\fbinv collected with  the CMS detector at the LHC.
Higgs bosons produced in
gluon-gluon fusion or in association with a $\cPZ$ boson are investigated.
Models including Higgs boson decays into a gravitino and a neutralino
or a pair of neutralinos, followed by the neutralino decay to
a gravitino and a photon, are tested. The measurements for the selected events in data are consistent
with the background only hypothesis, and the results are interpreted as limits on the product of cross sections
and branching fractions. Assuming a standard model Higgs production cross-section, a 95\% CL
upper limit is set on the branching fraction of a 125\GeV Higgs boson
decaying into undetectable particles and one or two isolated photons as a
function of the neutralino mass. For neutralino masses from 1 to 120\GeV an upper limit in the range of 7 to 13\% is obtained.
Further results are given as a function of the neutralino lifetime, and also for a range of Higgs boson masses.

\begin{acknowledgments}
We congratulate our colleagues in the CERN accelerator departments for the excellent performance of the LHC and thank the technical and administrative staffs at CERN and at other CMS institutes for their contributions to the success of the CMS effort. In addition, we gratefully acknowledge the computing centers and personnel of the Worldwide LHC Computing Grid for delivering so effectively the computing infrastructure essential to our analyses. Finally, we acknowledge the enduring support for the construction and operation of the LHC and the CMS detector provided by the following funding agencies: BMWFW and FWF (Austria); FNRS and FWO (Belgium); CNPq, CAPES, FAPERJ, and FAPESP (Brazil); MES (Bulgaria); CERN; CAS, MoST, and NSFC (China); COLCIENCIAS (Colombia); MSES and CSF (Croatia); RPF (Cyprus); MoER, ERC IUT and ERDF (Estonia); Academy of Finland, MEC, and HIP (Finland); CEA and CNRS/IN2P3 (France); BMBF, DFG, and HGF (Germany); GSRT (Greece); OTKA and NIH (Hungary); DAE and DST (India); IPM (Iran); SFI (Ireland); INFN (Italy); MSIP and NRF (Republic of Korea); LAS (Lithuania); MOE and UM (Malaysia); CINVESTAV, CONACYT, SEP, and UASLP-FAI (Mexico); MBIE (New Zealand); PAEC (Pakistan); MSHE and NSC (Poland); FCT (Portugal); JINR (Dubna); MON, RosAtom, RAS and RFBR (Russia); MESTD (Serbia); SEIDI and CPAN (Spain); Swiss Funding Agencies (Switzerland); MST (Taipei); ThEPCenter, IPST, STAR and NSTDA (Thailand); TUBITAK and TAEK (Turkey); NASU and SFFR (Ukraine); STFC (United Kingdom); DOE and NSF (USA).

Individuals have received support from the Marie-Curie program and the European Research Council and EPLANET (European Union); the Leventis Foundation; the A. P. Sloan Foundation; the Alexander von Humboldt Foundation; the Belgian Federal Science Policy Office; the Fonds pour la Formation \`a la Recherche dans l'Industrie et dans l'Agriculture (FRIA-Belgium); the Agentschap voor Innovatie door Wetenschap en Technologie (IWT-Belgium); the Ministry of Education, Youth and Sports (MEYS) of the Czech Republic; the Council of Science and Industrial Research, India; the HOMING PLUS program of the Foundation for Polish Science, cofinanced from European Union, Regional Development Fund; the Compagnia di San Paolo (Torino); the Consorzio per la Fisica (Trieste); MIUR project 20108T4XTM (Italy); the Thalis and Aristeia programs cofinanced by EU-ESF and the Greek NSRF; the National Priorities Research Program by Qatar National Research Fund; and Rachadapisek Sompot Fund for Postdoctoral Fellowship, Chulalongkorn University (Thailand).
\end{acknowledgments}

\bibliography{auto_generated}

\providecommand{\href}[2]{#2}\begingroup\raggedright\begin{thebibliography}{10}%
\makeatletter
\providecommand{\hrefCMSnoop }[0]{\@secondoftwo}%
\makeatother
\providecommand{\doi}{\texttt{doi:}\begingroup \urlstyle{tt}\Url}

\bibitem{AtlasPaperCombination}
\hrefCMSnoop {}{{ATLAS Collaboration}, ``{Observation of a new particle in the
  search for the Standard Model Higgs boson with the ATLAS detector at the
  LHC}'',} \textit{ Phys. Lett. B} \textbf{ 716} (2012) 1,
  \href{http://dx.doi.org/10.1016/j.physletb.2012.08.020}{\doi{10.1016/j.physletb.2012.08.020}},
  \href{http://www.arXiv.org/abs/1207.7214}{\texttt{arXiv:1207.7214}}.

\bibitem{CMSPaperCombination}
\hrefCMSnoop {}{{CMS Collaboration}, ``{Observation of a new boson at a mass of
  125 GeV with the CMS experiment at the LHC}'',} \textit{ Phys. Lett. B}
  \textbf{ 716} (2012) 30,
  \href{http://dx.doi.org/10.1016/j.physletb.2012.08.021}{\doi{10.1016/j.physletb.2012.08.021}},
  \href{http://www.arXiv.org/abs/1207.7235}{\texttt{arXiv:1207.7235}}.

\bibitem{CMSPaperCombination2}
\hrefCMSnoop {}{{CMS Collaboration}, ``{Observation of a new boson with mass
  near 125 GeV in pp collisions at $\sqrt{s}$ = 7 and 8 TeV}'',} \textit{ JHEP}
  \textbf{ 06} (2013) 081,
  \href{http://dx.doi.org/10.1007/JHEP06(2013)081}{\doi{10.1007/JHEP06(2013)081}},
  \href{http://www.arXiv.org/abs/1303.4571}{\texttt{arXiv:1303.4571}}.

\bibitem{Khachatryan:2014jba}
\hrefCMSnoop {}{{CMS Collaboration}, ``{Precise determination of the mass of
  the Higgs boson and tests of compatibility of its couplings with the standard
  model predictions using proton collisions at 7 and 8 $\,\text {TeV}$}'',}
  \textit{ Eur. Phys. J. C} \textbf{ 75} (2015), no.~5, 212,
  \href{http://dx.doi.org/10.1140/epjc/s10052-015-3351-7}{\doi{10.1140/epjc/s10052-015-3351-7}},
  \href{http://www.arXiv.org/abs/1412.8662}{\texttt{arXiv:1412.8662}}.

\bibitem{Khachatryan:2014kca}
\hrefCMSnoop {}{{CMS Collaboration}, ``{Constraints on the spin-parity and
  anomalous HVV couplings of the Higgs boson in proton collisions at 7 and 8
  TeV}'',} \textit{ Phys. Rev.} \textbf{ D92} (2015), no.~1, 012004,
  \href{http://dx.doi.org/10.1103/PhysRevD.92.012004}{\doi{10.1103/PhysRevD.92.012004}},
  \href{http://www.arXiv.org/abs/1411.3441}{\texttt{arXiv:1411.3441}}.

\bibitem{LHCHXSWG1}
\hrefCMSnoop {}{{LHC Higgs Cross Section Working Group}, S.~Dittmaier {et~al.},
  ``{Handbook of LHC Higgs Cross Sections: 1. Inclusive Observables}'',} CERN
  Report CERN-2011-002, 2011.
\newblock
  \href{http://dx.doi.org/10.5170/CERN-2011-002}{\doi{10.5170/CERN-2011-002}},
  \href{http://www.arXiv.org/abs/1101.0593}{\texttt{arXiv:1101.0593}}.

\bibitem{ZHTheory}
D.~Ghosh\hrefCMSnoop {}{ {et~al.}, ``{Looking for an invisible Higgs signal at
  the LHC}'',} \textit{ Phys. Lett. B} \textbf{ 725} (2012) 344,
  \href{http://dx.doi.org/10.1016/j.physletb.2013.07.042}{\doi{10.1016/j.physletb.2013.07.042}},
  \href{http://www.arXiv.org/abs/1211.7015}{\texttt{arXiv:1211.7015}}.

\bibitem{Martin:1999qf}
\hrefCMSnoop {}{S.~P. Martin and J.~D. Wells, ``{Motivation and detectability
  of an invisibly decaying Higgs boson at the Fermilab Tevatron}'',} \textit{
  Phys. Rev. D} \textbf{ 60} (1999) 035006,
  \href{http://dx.doi.org/10.1103/PhysRevD.60.035006}{\doi{10.1103/PhysRevD.60.035006}},
\href{http://www.arXiv.org/abs/hep-ph/9903259}{\texttt{arXiv:hep-ph/9903259}}.

\bibitem{Bai:2011wz}
\hrefCMSnoop {}{Y.~Bai, P.~Draper, and J.~Shelton, ``{Measuring the invisible
  Higgs width at the 7 and 8 TeV LHC}'',} \textit{ JHEP} \textbf{ 07} (2012)
  192,
  \href{http://dx.doi.org/10.1007/JHEP07(2012)192}{\doi{10.1007/JHEP07(2012)192}},
\href{http://www.arXiv.org/abs/1112.4496}{\texttt{arXiv:1112.4496}}.

\bibitem{Gori}
\hrefCMSnoop {}{{D. Curtin {\it et. al.}}, ``{Exotic decays of the 125 GeV
  Higgs boson}'',} \textit{ Phys. Rev. D} \textbf{ 90} (2014) 075004,
  \href{http://dx.doi.org/10.1103/PhysRevD.90.075004}{\doi{10.1103/PhysRevD.90.075004}},
\href{http://www.arXiv.org/abs/1312.4992}{\texttt{arXiv:1312.4992}}.

\bibitem{Djouadi1997243}
\hrefCMSnoop {}{A.~Djouadi and M.~Drees, ``Higgs boson decays into light
  gravitinos'',} \textit{ Phys. Lett. B} \textbf{ 407} (1997) 243,
  \href{http://dx.doi.org/10.1016/S0370-2693(97)00670-9}{\doi{10.1016/S0370-2693(97)00670-9}}.

\bibitem{Petersson:2012dp}
\hrefCMSnoop {}{C.~Petersson, A.~Romagnoni, and R.~Torre, ``{Higgs decay with
  monophoton + $\ensuremath{\not\!\!E_T}$ signature from low scale
  supersymmetry breaking}'',} \textit{ JHEP} \textbf{ 10} (2012) 016,
  \href{http://dx.doi.org/10.1007/JHEP10(2012)016}{\doi{10.1007/JHEP10(2012)016}},
  \href{http://www.arXiv.org/abs/1203.4563}{\texttt{arXiv:1203.4563}}.

\bibitem{cms-hig-13-006}
\hrefCMSnoop {}{{CMS Collaboration}, ``Search for a Higgs boson decaying into a
  {\Z} and a photon in {\Pp\Pp} collisions at $\sqrt{s}$ = 7 and 8{\TeV}'',}
  \textit{ Phys. Lett. B} \textbf{ 726} (2013) 587,
  \href{http://dx.doi.org/10.1016/j.physletb.2013.09.057}{\doi{10.1016/j.physletb.2013.09.057}}.

\bibitem{atlas-hig-14-zgamma}
\hrefCMSnoop {}{{ATLAS Collaboration}, ``Search for {H}iggs boson decays to a
  photon and a {Z} boson in pp collisions at $\sqrt{s}$ = 7 and 8 {TeV} with
  the {ATLAS} detector'',} \textit{ Phys. Lett. B} \textbf{ 732} (2014) 8,
  \href{http://dx.doi.org/10.1016/j.physletb.2014.03.015}{\doi{10.1016/j.physletb.2014.03.015}},
  \href{http://www.arXiv.org/abs/1402.3051}{\texttt{arXiv:1402.3051}}.

\bibitem{CMSdetector}
\hrefCMSnoop {}{{CMS Collaboration}, ``The {CMS} experiment at the {CERN
  LHC}'',} \textit{ JINST} \textbf{ 3} (2008) S08004,
  \href{http://dx.doi.org/10.1088/1748-0221/3/08/S08004}{\doi{10.1088/1748-0221/3/08/S08004}}.

\bibitem{CMS-PAS-PFT-09-001}
\href {http://cdsweb.cern.ch/record/1194487}{{CMS Collaboration},
  ``Particle--Flow Event Reconstruction in {CMS} and Performance for Jets,
  Taus, and {\MET}'',} CMS Physics Analysis Summary CMS-PAS-PFT-09-001, 2009.

\bibitem{CMS-PAS-PFT-10-001}
\href {http://cdsweb.cern.ch/record/1247373}{{CMS Collaboration},
  ``Commissioning of the particle-flow event reconstruction with the first
  {LHC} collisions recorded in the {CMS} detector'',} CMS Physics Analysis
  Summary CMS-PAS-PFT-10-001, 2010.

\bibitem{antikt}
\hrefCMSnoop {}{{M. Cacciari and G. P. Salam and G. Soyez}, ``{The anti-$k_t$
  jet clustering algorithm}'',} \textit{ JHEP} \textbf{ 04} (2008) 063,
  \href{http://dx.doi.org/10.1088/1126-6708/2008/04/063}{\doi{10.1088/1126-6708/2008/04/063}},
  \href{http://www.arXiv.org/abs/0802.1189}{\texttt{arXiv:0802.1189}}.

\bibitem{Cacciari:fastjet1}
\hrefCMSnoop {}{M.~Cacciari, G.~P. Salam, and G.~Soyez, ``FastJet user
  manual'',} \textit{ Eur. Phys. J. C} \textbf{ 72} (2012) 1896,
  \href{http://dx.doi.org/10.1140/epjc/s10052-012-1896-2}{\doi{10.1140/epjc/s10052-012-1896-2}},
  \href{http://www.arXiv.org/abs/hep-ph/1111.6097}{\texttt{arXiv:hep-ph/1111.6097}}.

\bibitem{Cacciari:fastjet2}
\hrefCMSnoop {}{M.~Cacciari and G.~P. Salam, ``{Dispelling the $N^{3}$ myth for
  the $k_t$ jet-finder}'',} \textit{ Phys. Lett. B} \textbf{ 641} (2006) 57,
  \href{http://dx.doi.org/10.1016/j.physletb.2006.08.037}{\doi{10.1016/j.physletb.2006.08.037}},
  \href{http://www.arXiv.org/abs/hep-ph/0512210}{\texttt{arXiv:hep-ph/0512210}}.

\bibitem{jetIdPAS}
\href {http://cdsweb.cern.ch/record/1581583}{{CMS Collaboration}, ``Pileup Jet
  Identification'',} CMS Physics Analysis Summary CMS-PAS-JME-13-005, 2013.

\bibitem{madgraph}
J.~Alwall\hrefCMSnoop {}{ {et~al.}, ``{MadGraph/MadEvent v4}: the new web
  generation'',} \textit{ JHEP} \textbf{ 09} (2007) 028,
  \href{http://dx.doi.org/10.1088/1126-6708/2007/09/028}{\doi{10.1088/1126-6708/2007/09/028}}.

\bibitem{ggww}
\hrefCMSnoop {}{T.~Binoth, M.~Ciccolini, N.~Kauer, and M.~Kr{\"a}mer,
  ``Gluon-induced {$W$}-boson pair production at the {LHC}'',} \textit{ JHEP}
  \textbf{ 12} (2006) 046,
  \href{http://dx.doi.org/10.1088/1126-6708/2006/12/046}{\doi{10.1088/1126-6708/2006/12/046}}.

\bibitem{pythia}
\hrefCMSnoop {}{T.~Sj{\"o}strand, S.~Mrenna, and P.~Skands, ``PYTHIA 6.4
  physics and manual'',} \textit{ JHEP} \textbf{ 05} (2006) 026,
  \href{http://dx.doi.org/10.1088/1126-6708/2006/05/026}{\doi{10.1088/1126-6708/2006/05/026}}.

\bibitem{cteq66}
H.-L. Lai\hrefCMSnoop {}{ {et~al.}, ``Uncertainty induced by {QCD} coupling in
  the {CTEQ} global analysis of parton distributions'',} \textit{ Phys. Rev. D}
  \textbf{ 82} (2010) 054021,
  \href{http://dx.doi.org/10.1103/PhysRevD.82.054021}{\doi{10.1103/PhysRevD.82.054021}},
  \href{http://www.arXiv.org/abs/1004.4624}{\texttt{arXiv:1004.4624}}.

\bibitem{Lai:2010vv}
H.-L. Lai\hrefCMSnoop {}{ {et~al.}, ``New parton distributions for collider
  physics'',} \textit{ Phys. Rev. D} \textbf{ 82} (2010) 074024,
  \href{http://dx.doi.org/10.1103/PhysRevD.82.074024}{\doi{10.1103/PhysRevD.82.074024}},
  \href{http://www.arXiv.org/abs/1007.2241}{\texttt{arXiv:1007.2241}}.

\bibitem{Agostinelli:2002hh}
\hrefCMSnoop {}{{GEANT4} Collaboration, ``{GEANT4}---a simulation toolkit'',}
  \textit{ Nucl. Instrum. Meth. A} \textbf{ 506} (2003) 250,
\href{http://dx.doi.org/10.1016/S0168-9002(03)01368-8}{\doi{10.1016/S0168-9002(03)01368-8}}.

\bibitem{Khachatryan:2015iwa}
\hrefCMSnoop {}{{CMS Collaboration}, ``Performance of photon reconstruction and
  identification with the CMS detector in proton-proton collisions at
  {$\sqrt{s} = 8$\TeV}'',} \textit{ JINST} \textbf{ 10} (2015), no.~08, P08010,
  \href{http://dx.doi.org/10.1088/1748-0221/10/08/P08010}{\doi{10.1088/1748-0221/10/08/P08010}},
  \href{http://www.arXiv.org/abs/1502.02702}{\texttt{arXiv:1502.02702}}.

\bibitem{Khachatryan:2014rwa}
\hrefCMSnoop {}{{CMS Collaboration}, ``{Search for new phenomena in monophoton
  final states in proton-proton collisions at $\sqrt{s}$ = 8 TeV}'',} (2014).
  \href{http://www.arXiv.org/abs/1410.8812}{\texttt{arXiv:1410.8812}}.
Submitted to \textit{Phys. Lett. B}.

\bibitem{Khachatryan:2015hwa}
\hrefCMSnoop {}{{CMS Collaboration}, ``{Performance of electron reconstruction
  and selection with the {CMS} detector in proton-proton collisions at
  {$\sqrt{s}=8\TeV$}}'',} \textit{ JINST} \textbf{ 10} (2015), no.~06, P06005,
  \href{http://dx.doi.org/10.1088/1748-0221/10/06/P06005}{\doi{10.1088/1748-0221/10/06/P06005}},
  \href{http://www.arXiv.org/abs/1502.02701}{\texttt{arXiv:1502.02701}}.

\bibitem{Chatrchyan:2011tn}
\hrefCMSnoop {}{{CMS Collaboration}, ``Missing transverse energy performance of
  the {CMS} detector'',} \textit{ JINST} \textbf{ 6} (2011) P09001,
  \href{http://dx.doi.org/10.1088/1748-0221/6/09/P09001}{\doi{10.1088/1748-0221/6/09/P09001}}.

\bibitem{hig13023}
\hrefCMSnoop {}{{CMS Collaboration}, ``Measurement of {H}iggs boson production
  and properties in the {WW} decay channel with leptonic final states'',}
  \textit{ JHEP} \textbf{ 01} (2014) 096,
  \href{http://dx.doi.org/10.1007/JHEP01(2014)096}{\doi{10.1007/JHEP01(2014)096}}.

\bibitem{hig13001}
\hrefCMSnoop {}{{CMS Collaboration}, ``Observation of the diphoton decay of the
  {H}iggs boson and measurement of its properties'',} \textit{ Eur. Phys. J. C}
  \textbf{ 74} (2014) 3076,
  \href{http://dx.doi.org/10.1140/epjc/s10052-014-3076-z}{\doi{10.1140/epjc/s10052-014-3076-z}},
  \href{http://www.arXiv.org/abs/1407.0558}{\texttt{arXiv:1407.0558}}.

\bibitem{btag}
\hrefCMSnoop {}{{CMS Collaboration}, ``Identification of b-quark jets with the
  {CMS} experiment'',} \textit{ J. Instrum.} \textbf{ 8} (2012) P04013,
  \href{http://dx.doi.org/10.1088/1748-0221/8/04/P04013}{\doi{10.1088/1748-0221/8/04/P04013}}.

\bibitem{CMS:2011aa}
\hrefCMSnoop {}{{CMS Collaboration}, ``Measurement of the inclusive {W} and {Z}
  production cross sections in pp collisions at {$\sqrt{s}=7\TeV$} with the
  {CMS} experiment'',} \textit{ JHEP} \textbf{ 10} (2011) 132,
  \href{http://dx.doi.org/10.1007/JHEP10(2011)132}{\doi{10.1007/JHEP10(2011)132}}.

\bibitem{wzxs}
\hrefCMSnoop {}{{CMS Collaboration}, ``Measurements of inclusive {W} and {Z}
  cross sections in pp collisions at $\sqrt{s}$ = 7 {TeV}'',} \textit{ J. High
  Energy Phys.} \textbf{ 01} (2011) 080,
  \href{http://dx.doi.org/10.1007/JHEP01(2011)080}{\doi{10.1007/JHEP01(2011)080}}.

\bibitem{CMSinvSearch}
\hrefCMSnoop {}{{CMS Collaboration}, ``Search for invisible decays of {H}iggs
  bosons in the vector boson fusion and associated {ZH} production modes'',}
  \textit{ Eur. Phys. J. C} \textbf{ 74} (2014) 2980,
  \href{http://dx.doi.org/10.1140/epjc/s10052-014-2980-6}{\doi{10.1140/epjc/s10052-014-2980-6}}.

\bibitem{CMS:2013gfa}
\href {http://cdsweb.cern.ch/record/1598864}{{CMS Collaboration}, ``CMS
  Luminosity Based on Pixel Cluster Counting - Summer 2013 Update'',} CMS
  Physics Analysis Summary CMS-PAS-LUM-13-001, 2013.

\bibitem{Botje:2011sn}
M.~Botje\hrefCMSnoop {}{ {et~al.}, ``{The PDF4LHC Working Group Interim
  Recommendations}'',} (2011).
\href{http://www.arXiv.org/abs/1101.0538}{\texttt{arXiv:1101.0538}}.

\bibitem{Alekhin:2011sk}
\hrefCMSnoop {}{S.~Alekhin {et~al.}, ``{The PDF4LHC Working Group Interim
  Report}'',} (2011).
  \href{http://www.arXiv.org/abs/1101.0536}{\texttt{arXiv:1101.0536}}.

\bibitem{Martin:2009iq}
\hrefCMSnoop {}{A.~D. Martin, W.~J. Stirling, R.~S. Thorne, and G.~Watt,
  ``Parton distributions for the {LHC}'',} \textit{ Eur. Phys. J. C} \textbf{
  63} (2009) 189,
  \href{http://dx.doi.org/10.1140/epjc/s10052-009-1072-5}{\doi{10.1140/epjc/s10052-009-1072-5}},
  \href{http://www.arXiv.org/abs/0901.0002}{\texttt{arXiv:0901.0002}}.

\bibitem{Ball:2011mu}
\hrefCMSnoop {}{{NNPDF} Collaboration, ``Impact of heavy quark masses on parton
  distributions and {LHC} phenomenology'',} \textit{ Nucl. Phys. B} \textbf{
  849} (2011)
  \href{http://dx.doi.org/10.1016/j.nuclphysb.2011.03.021}{\doi{10.1016/j.nuclphysb.2011.03.021}},
  \href{http://www.arXiv.org/abs/1101.1300}{\texttt{arXiv:1101.1300}}.

\bibitem{MCFM}
\hrefCMSnoop {}{J.~M. Campbell and R.~K. Ellis, ``{MCFM} for the {Tevatron} and
  the {LHC}'',} \textit{ Nucl. Phys. Proc. Suppl.} (2010) 10,
  \href{http://dx.doi.org/10.1016/j.nuclphysbps.2010.08.011}{\doi{10.1016/j.nuclphysbps.2010.08.011}}.

\bibitem{LHC-HCG-Report}
\href {http://cdsweb.cern.ch/record/1379837}{{ATLAS and CMS Collaborations, LHC
  Higgs Combination Group}, ``Procedure for the {LHC} {H}iggs boson search
  combination in {S}ummer 2011'',} Technical Report ATL-PHYS-PUB 2011-11, CMS
  NOTE 2011/005, 2011.

\bibitem{Read1}
\hrefCMSnoop {}{A.~L. Read, ``Presentation of search results: the {$CL_s$}
  technique'',} \textit{ J. Phys. G} \textbf{ 28} (2002) 2693,
\href{http://dx.doi.org/10.1088/0954-3899/28/10/313}{\doi{10.1088/0954-3899/28/10/313}}.

\bibitem{junkcls}
\hrefCMSnoop {}{T.~Junk, ``{Confidence level computation for combining searches
  with small statistics}'',} \textit{ Nucl. Instrum. Meth. A} \textbf{ 434}
  (1999) 435,
  \href{http://dx.doi.org/10.1016/S0168-9002(99)00498-2}{\doi{10.1016/S0168-9002(99)00498-2}}.

\end{thebibliography}\endgroup

\cleardoublepage \appendix\section{The CMS Collaboration \label{app:collab}}\begin{sloppypar}\hyphenpenalty=5000\widowpenalty=500\clubpenalty=5000\textbf{Yerevan Physics Institute,  Yerevan,  Armenia}\\*[0pt]
V.~Khachatryan, A.M.~Sirunyan, A.~Tumasyan
\vskip\cmsinstskip
\textbf{Institut f\"{u}r Hochenergiephysik der OeAW,  Wien,  Austria}\\*[0pt]
W.~Adam, E.~Asilar, T.~Bergauer, J.~Brandstetter, E.~Brondolin, M.~Dragicevic, J.~Er\"{o}, M.~Flechl, M.~Friedl, R.~Fr\"{u}hwirth\cmsAuthorMark{1}, V.M.~Ghete, C.~Hartl, N.~H\"{o}rmann, J.~Hrubec, M.~Jeitler\cmsAuthorMark{1}, V.~Kn\"{u}nz, A.~K\"{o}nig, M.~Krammer\cmsAuthorMark{1}, I.~Kr\"{a}tschmer, D.~Liko, T.~Matsushita, I.~Mikulec, D.~Rabady\cmsAuthorMark{2}, B.~Rahbaran, H.~Rohringer, J.~Schieck\cmsAuthorMark{1}, R.~Sch\"{o}fbeck, J.~Strauss, W.~Treberer-Treberspurg, W.~Waltenberger, C.-E.~Wulz\cmsAuthorMark{1}
% ...
\vskip\cmsinstskip
% ...
1:~~Also at Vienna University of Technology, Vienna, Austria\\
2:~~Also at CERN, European Organization for Nuclear Research, Geneva, Switzerland\\
% ...

\end{sloppypar}
\end{document}